\documentclass{elsarticle}

\usepackage[reviewcopy]{adndt}
\usepackage{longtable}

\usepackage{amsmath}

\usepackage{amssymb}

\biboptions{square,sort&compress}
\bibpunct[]{[}{]}{,}{n}{}{;}
\begin{document}

\begin{frontmatter}

\journal{Atomic Data and Nuclear Data Tables}


\title{Discovery of the actinium, thorium, protactinium, and uranium isotopes}

\author{C. Fry}
\author{M. Thoennessen\corref{cor1}}\ead{thoennessen@nscl.msu.edu}

 \cortext[cor1]{Corresponding author.}

 \address{National Superconducting Cyclotron Laboratory and \\ Department of Physics and Astronomy, Michigan State University, \\ East Lansing, MI 48824, USA}

\begin{abstract}
Currently, 31 actinium, 31 thorium, 28 protactinium, and 23 uranium isotopes have so far been observed; the discovery of these isotopes is discussed. For each isotope a brief summary of the first refereed publication, including the production and identification method, is presented.
\end{abstract}

\end{frontmatter}





\newpage
\tableofcontents
\listofDtables

\vskip5pc

\section{Introduction}\label{s:intro}

The discovery of actinium, thorium, protactinium, and uranium isotopes is discussed as part of the series summarizing the discovery of isotopes, beginning with the cerium isotopes in 2009 \cite{2009Gin01}. Guidelines for assigning credit for discovery are (1) clear identification, either through decay-curves and relationships to other known isotopes, particle or $\gamma$-ray spectra, or unique mass and Z-identification, and (2) publication of the discovery in a refereed journal. The authors and year of the first publication, the laboratory where the isotopes were produced as well as the production and identification methods are discussed. When appropriate, references to conference proceedings, internal reports, and theses are included. When a discovery includes a half-life measurement the measured value is compared to the currently adopted value taken from the NUBASE evaluation \cite{2003Aud01} which is based on the ENSDF database \cite{2008ENS01}. In cases where the reported half-life differed significantly from the adopted half-life (up to approximately a factor of two), we searched the subsequent literature for indications that the measurement was erroneous. If that was not the case we credited the authors with the discovery in spite of the inaccurate half-life. All reported half-lives inconsistent with the presently adopted half-life for the ground state were compared to isomer half-lives and accepted as discoveries if appropriate following the criterium described above.

The first criterium is not clear cut and in many instances debatable. Within the scope of the present project it is not possible to scrutinize each paper for the accuracy of the experimental data as is done for the discovery of elements \cite{1991IUP01}. In some cases an initial tentative assignment is not specifically confirmed in later papers and the first assignment is tacitly accepted by the community. The readers are encouraged to contact the authors if they disagree with an assignment because they are aware of an earlier paper or if they found evidence that the data of the chosen paper were incorrect.

The discovery of several isotopes has only been reported in conference proceedings which are not accepted according to the second criterium. One example from fragmentation experiments why publications in conference proceedings should not be considered is $^{118}$Tc and $^{120}$Ru which had been reported as being discovered in a conference proceeding \cite{1996Cza01} but not in the subsequent refereed publication \cite{1997Ber01}.

The initial literature search was performed using the databases ENSDF \cite{2008ENS01} and NSR \cite{2008NSR01} of the National Nuclear Data Center at Brookhaven National Laboratory. These databases are complete and reliable back to the early 1960's. For earlier references, several editions of the Table of Isotopes were used \cite{1940Liv01,1944Sea01,1948Sea01,1953Hol02,1958Str01,1967Led01}.
For the isotopes of the radioactive decay chains several books and articles were consulted, for example, the 1908 edition of ``Gmelin-Kraut's Handbuch der anorganischen Chemie'' \cite{1908Fri01}, Soddy's 1911 book ``The chemistry of the radio-elements'' \cite{1911Sod01}, the 1913 edition of Rutherford's book ``Radioactive substances and their radiations'' \cite{1913Rut01}, and the 1933 article by Mary Elvira Weeks ``The discovery of the elements. XIX. The radioactive elements'' published in the Journal of Chemical Education \cite{1933Wee01}. In addition, the wikipedia page on the radioactive decay chains was a good starting point \cite{2011wik02}.

The isotopes within the radioactive decay chains were treated differently. Their decay properties were largely measured before the concept of isotopes was established. Thus we gave credit to the first observation and identification of a specific activity, even when it was only later placed properly within in the decay chain.

\begin{figure}
	\centering
	\includegraphics[scale=.75]{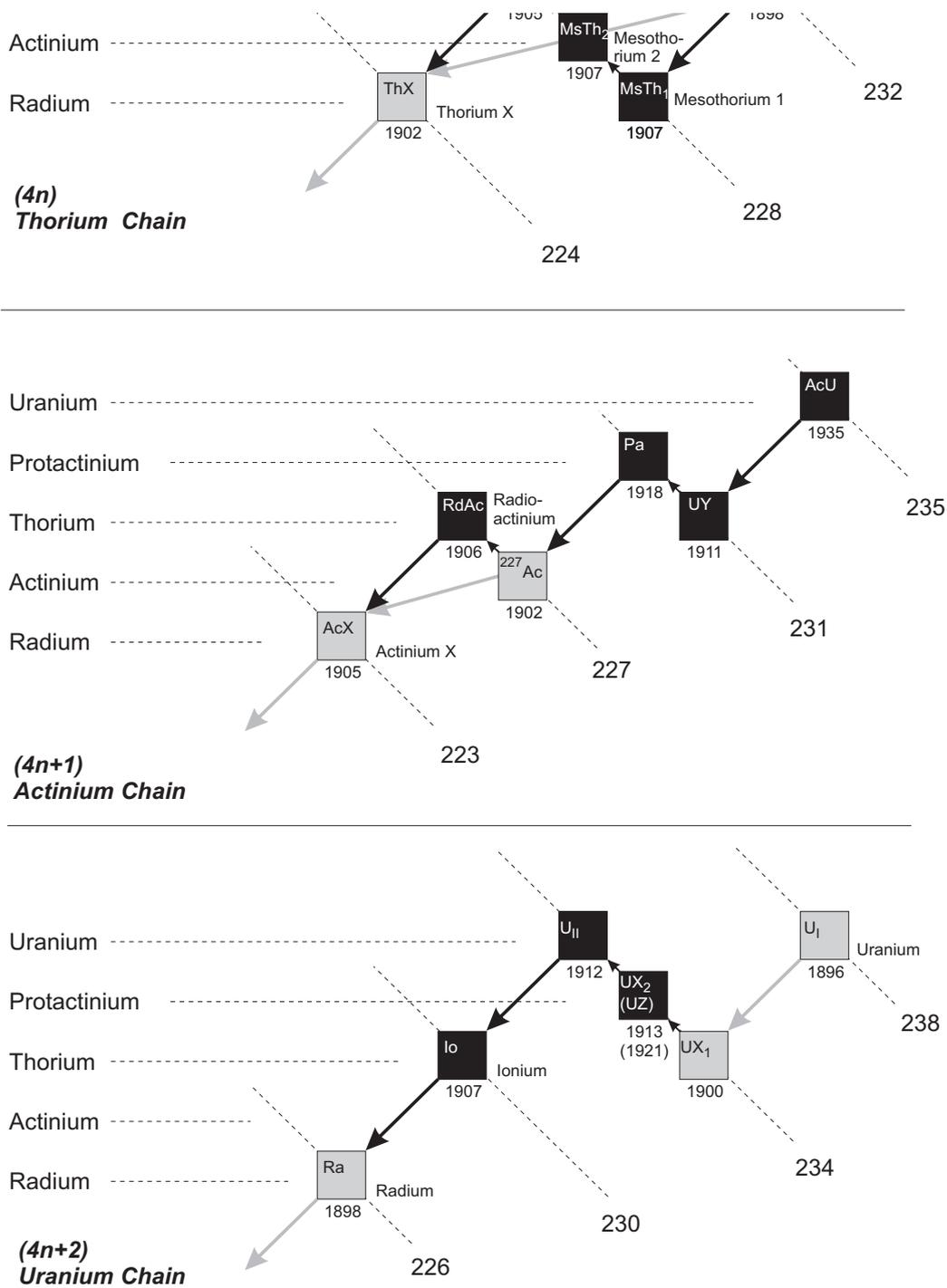}
	\caption{Original nomenclature of radium, actinium, thorium, protactinium, and uranium isotopes within the three natural occurring radioactive decay series. The grey squares connected by the grey arrows depict the activities labeled by Rutherford in his Bakerian lecture \cite{1905Rut01}. The black squares correspond to radioactive substances discovered later. }
\label{f:chain}
\end{figure}

Figure \ref{f:chain} summarizes the isotopes of the three natural occurring radioactive decay series with their original nomenclature. This notation of the original substances introduced by Rutherford during his Bakerian lecture presented on May 19$^{th}$ 1904 and published a year later \cite{1905Rut01} are shown by grey squares and connected by the grey arrows representing $\alpha$ and $\beta$ decay. The uranium and radium decay chains were later shown to be connected and the decay from actinium to actinium X and from thorium to thorium X were more complex. These isotopes are shown as black squares with the corresponding decays shown by black arrows.


\section{$^{206-236}$Ac}\vspace{0.0cm}

The discovery of the element actinium is generally credited to Debierne who described the observation of a new radioactive substance which he named actinium in two papers in 1899 and 1900 \cite{1899Deb01,1900Deb01}. However, in a comprehensive review of the literature Kirby argued in 1971 that actinium discovered by Giesel in 1902 \cite{1902Gie01}. Giesel had named his new substance emanium \cite{1904Gie01}.

\begin{figure}
	\centering
	\includegraphics[scale=.7]{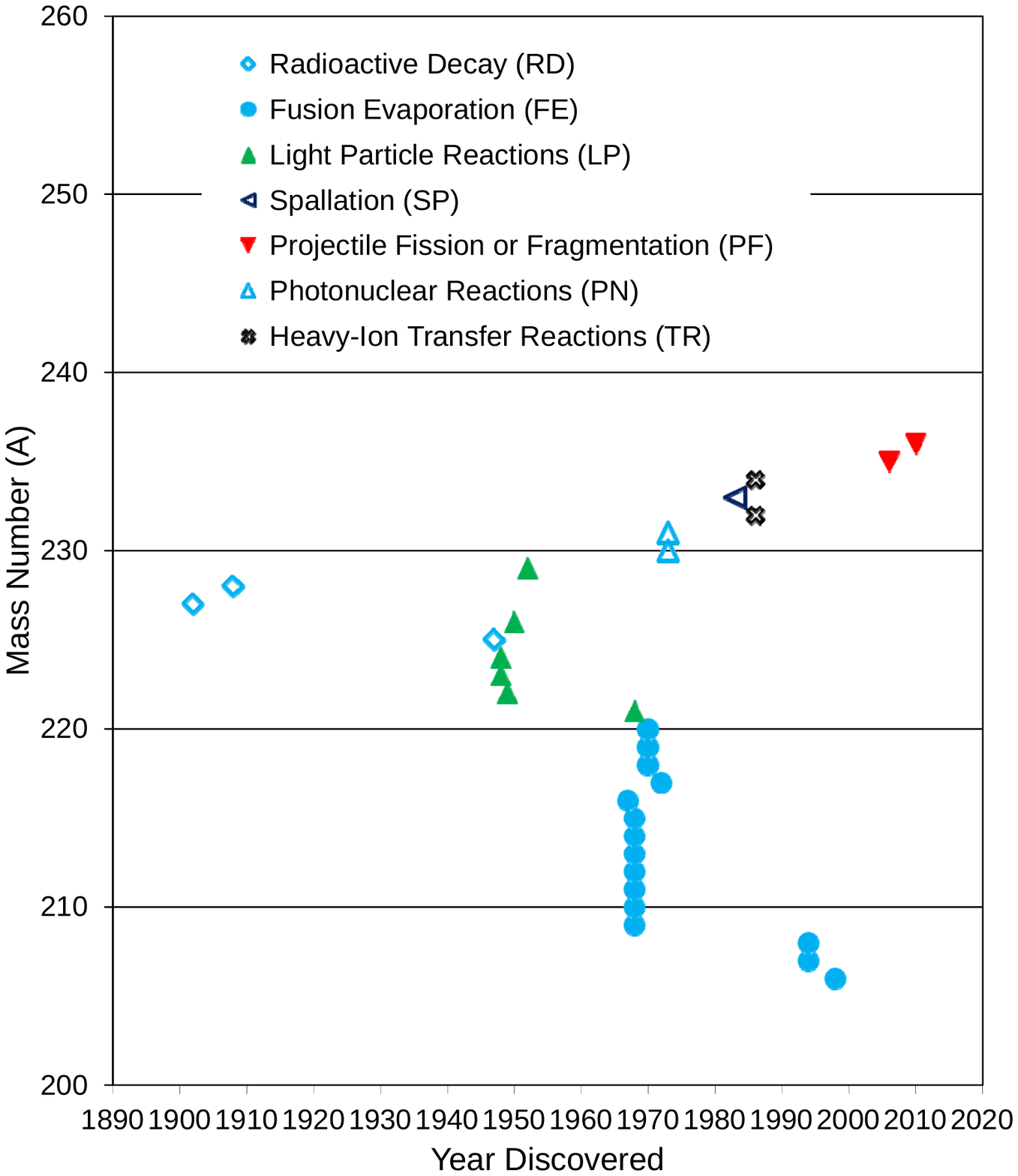}
	\caption{Actinium isotopes as a function of time when they were discovered. The different production methods are indicated.}
\label{f:year-actinium}
\end{figure}

Thirty-one actinium isotopes from A = 206--236 have been discovered so far and according to the HFB-14 model \cite{2007Gor01} about 60 additional actinium isotopes could exist. Figure \ref{f:year-actinium} summarizes the year of first discovery for all actinium isotopes identified by the method of discovery: radioactive decay (RD), fusion evaporation reactions (FE), light-particle reactions (LP), projectile fragmentation (PF), spallation (SP), photo-nuclear (PN), and heavy-ion transfer reactions (TR). In the following, the discovery of each actinium isotope is discussed in detail and a summary is presented in Table 1.

\subsection*{$^{206}$Ac}
Eskola et al.\ discovered $^{206}$Ac and published the results in the 1998 paper ``$\alpha$ decay of the new isotope $^{206}$Ac'' \cite{1998Esk01}. A 5.5~MeV/nucleon $^{36}$Ar beam from the Jyv\"askyl\"a K-130 heavy-ion cyclotron bombarded a $^{175}$Lu target to form $^{206}$Ac in (5n) fusion-evaporation reactions. Residues were separated with the gas-filled separator RITU and implanted in a position sensitive passivated planar silicon detector which also detected subsequent $\alpha$ decay. ``$^{206}$Ac was found to have two $\alpha$ particle emitting isomeric levels with half-lives of (22$^{+9}_{-5}$)~ms and (33$^{+22}_{-9}$)~ms, and with $\alpha$ particle energies of (7790$\pm$30)~keV and (7750$\pm$20)~keV, respectively.'' These half-lives correspond to the currently adopted values for the ground state and an isomeric state, respectively.

\subsection*{$^{207,208}$Ac}
The discovery of $^{207}$Ac and $^{208}$Ac was reported by Leino et al.\ in the 1994 paper ``Alpha decay of the new isotopes $^{207,208}$Ac'' \cite{1994Lei01}. A $^{175}$Lu target was bombarded with 5.2$-$5.6~MeV $^{40}$Ar beams from the Jyv\"askyl\"a K-130 heavy-ion cyclotron producing $^{207}$Ac and $^{208}$Ac in (8n) and (7n) fusion-evaporation reactions, respectively. Residues were separated with the gas-filled recoil separator RITU and implanted in a position sensitive PIPS detector which also recorded subsequent $\alpha$ decay. ``The alpha energy and half-life of $^{208}$Ac were determined to be (7572$\pm$15)~keV and (95$^{+24}_{-16}$)~ms, respectively. A new alpha line with a half-life of (25$^{+9}_{-5}$)~ms and an energy of (7758$\pm$20)~keV is assigned to the decay of an isomeric state in $^{208}$Ac. Another new activity with a half-life of (22$^{+40}_{-9}$)~ms and an alpha energy of (7712$\pm$25)~keV is assigned to $^{207}$Ac.'' The value for $^{208}$Ac is the currently adopted half-life and the quoted half-life for $^{207}$Ac agrees with the currently accepted value of 27($^{+11}_{-6}$)~ms.

\subsection*{$^{209-215}$Ac}
In 1968 Valli et al.\ reported the first observation of $^{209}$Ac, $^{210}$Ac, $^{211}$Ac, $^{212}$Ac, $^{213}$Ac, $^{214}$Ac, and $^{215}$Ac in the article ``On-line alpha spectroscopy of neutron-deficient actinium isotopes'' \cite{1968Val01}. The Berkeley heavy-ion linear accelerator was used to produce light actinium isotopes in the reactions $^{197}$Au($^{20}$Ne,xn), $^{203,205}$Tl($^{16}$O,xn), and $^{209}$Bi($^{12}$C,xn). Reaction products were deposited by helium flow onto a catcher foil which was then rotated in front of a Si(Au) surface barrier detector. ``Actinium-215: ...The shape of the excitation function for the 7.602-MeV $\alpha$ activity in the $^{203}$Tl+$^{16}$O case is particularly good evidence for the assignment to $^{215}$Ac... Actinium-214: ...The similarity of the curves for the peaks at 7.212, 7.080, and 7.000~MeV makes us conclude that they are associated with a single isotope. The maximum yield occurs at bombarding energies which are reasonable for the $^{214}$Ac assignment in the four reactions studied... Actinium-213 and Actinium-212: ...There is a peak in the $\alpha$ spectra in [the figures] which we attribute to $^{213}$Ac with an energy of 7.362~MeV, to $^{212}$Ac with an energy of 7.377~MeV, or to a mixture of both, depending on the reaction system and the beam energy... Actinium-211 and Actinium-210: ...In the $^{197}$Au+$^{20}$Ne system, when we prepared the activity with 124-MeV $^{20}$Ne ions, the $\alpha$ energy was 7.480~MeV. This energy we assign to $^{211}$Ac. When we used 145-MeV $^{20}$Ne ions, the $\alpha$ energy of the unknown peak was 7.462~MeV, which we assign to $^{210}$Ac... Actinium-209: In the $^{197}$Au+$^{20}$Ne series of experiments, we observed an $\alpha$ activity with 7.585-MeV energy and a 0.10$\pm$0.05-sec half-life, which we assign to $^{209}$Ac.'' The measured half-lives of 0.10(5)~s for $^{209}$Ac, 0.25(5)~s for $^{211}$Ac, and 0.80(5)~s for $^{213}$Ac agree with the presently adopted values of 0.092(11)~s, 0.21(3)~s, and 0.738(16)~s, respectively. The half-lives of 0.35(5)~s for $^{210}$Ac, 0.93(5)~s for $^{212}$Ac, 8.2(2)~s for $^{214}$Ac, and 0.17(1)~s for $^{215}$Ac correspond to the currently adopted values.

\subsection*{$^{216}$Ac}
Rotter et al.\ discovered $^{216}$Ac in 1967 and reported their results in the paper ``The new isotope Ac$^{216}$'' \cite{1967Rot01}. An 80~MeV $^{12}$C beam from the Dubna 1.5~m cyclotron bombarded bismuth targets producing $^{216}$Ac in (5n) fusion-evaporation reactions. Recoil nuclei were collected on an aluminum foil and $\alpha$-particle spectra were measured with a silicon surface barrier detector. ``When bismuth was bombarded with carbon ions of maximum energy a 9.14-MeV $\alpha$ line was discovered... Several measurements yielded (0.39$\pm$0.04)~msec as the half-life of this line... We may assign the observed activity to Ac$^{216}$ produced in the reaction Bi$^{209}$(C$^{12}$,5n)Ac$^{216}$.'' This half-life is close to the currently adopted value of 440(16)~$\mu$s.

\subsection*{$^{217}$Ac}
In the article ``In-beam alpha spectroscopy of N=128 isotones. Lifetimes of $^{216}$Ra and a new isotope $^{217}$Ac,'' Nomura et al.\ reported the observation of $^{217}$Ac in 1972 \cite{1972Nom01}. A $^{208}$Pb target was bombarded with a 91~MeV $^{14}$N beam from the RIKEN IPCR cyclotron forming $^{217}$Ac in (5n) fusion-evaporation reactions. Alpha-particle spectra and decay curves were measured with a surface barrier Si detector. ``Time distributions of the ground-state decay of $^{216}$Ra and $^{217}$Ac are shown in [the figure], from which half-lives of $^{216}$Ra and $^{217}$Ac have been determined of 0.18$\pm$0.03$~\mu$s and 0.10$\pm$0.01~$\mu$s, respectively.'' The quoted value for $^{217}$Ac is close to the currently adopted half-life of 69(4)~ns.

\subsection*{$^{218-220}$Ac}
In the 1970 article ``Production and decay properties of protactinium isotopes of mass 222 to 225 formed in heavy-ion reactions,'' Borggreen et al.\ identified $^{218}$Ac, $^{219}$Ac, and $^{220}$Ac \cite{1970Bor01}. The Berkeley heavy-ion linear accelerator (HILAC) was used to bombard $^{209}$Bi, $^{208}$Pb and $^{205}$Tl targets with $^{16}$O, $^{19}$F and $^{22}$Ne beams forming $^{222}$Pa, $^{223}$Pa, and $^{224}$Pa in (xn) fusion-evaporation reactions. $^{218}$Ac, $^{219}$Ac, and $^{220}$Ac were then populated by $\alpha$-decay. Recoil products were deposited by a helium gas stream on a metal surface located in front of a gold surface-barrier detector which recorded the subsequent $\alpha$ decay. ``The half-life of $^{220}$Ac was measured in an experiment in which a sample was collected during a 104-msec period and then, with the beam off, the $\alpha$ spectrum was recorded for 16 time periods of 20-msec each. The decay of individual peaks of $^{220}$Ac and of the 9.005-MeV peak of $^{216}$Fr were plotted separately. The most accurate value of the halflife (26.1$\pm$0.5~msec) came from the decay of the $^{216}$Fr peak as shown in [the figure]... The intensity of the 8.18-MeV $^{223}$Pa peak, the 8.66-MeV $^{219}$Ac peak, and the 9.365-MeV $^{215}$Fr peak was plotted versus the time elapsed from the time of beam interruption... The time plots corresponding to these data sortings revealed a one-component halflife of 7$\pm$2~$\mu$sec for $^{219}$Ac... The possible assignments of an $\alpha$ energy of this magnitude are extremely limited; the favored assignment is to $^{218}$Ac... $^{218}$Ac has a 0.27-$\mu$sec half-life, so that it appears in these spectra only by its continued replenishment by the decay of $^{222}$Pa.'' The half-lives of 7(2)~$\mu$s for $^{219}$Ac and 26.1(5)~ms for $^{220}$Ac agree with the presently adopted values of 11.8(15)~$\mu$s and 26.36(19)~ms, respectively. The half-life of 0.27(4)~$\mu$s for $^{218}$Ac differs significantly from the current value of 1.08(9)~$\mu$s, however, the $\alpha$-decay energy was correct and the relationship to the parent $^{222}$Pa was established.

\subsection*{$^{221}$Ac}
The first detection of $^{221}$Ac was reported in 1968 by Hahn et al.\ in ``New neptunium isotopes, $^{230}$Np and $^{229}$Np'' \cite{1968Hah01}. $^{233}$U was bombarded with 32$-$41.6~MeV protons from the Oak Ridge isochronous cyclotron forming $^{229}$Np in (p,5n) reactions. Reaction products were implanted on a catcher foil which was periodically rotated in front of a surface barrier Si(Au) detector which measured subsequent $\alpha$ decay. ``The $\alpha$-particle energies found for the $^{225}$Pa series are more precise than the previously available values: $^{225}$Pa, 7.25$\pm$0.02~MeV (new value); $^{221}$Ac, 7.63$\pm$0.02~MeV; $^{217}$Fr, 8.31$\pm$0.02~MeV and $^{213}$At, 9.06$\pm$0.02~MeV.'' The observation of $^{221}$Ac was not considered new, referring to an unpublished thesis \cite{1951Key01}. The currently accepted half-life is 52(2)~ms.

\subsection*{$^{222}$Ac}
Meinke et al.\ reported the observation of $^{222}$Ac in the 1949 paper ``Three additional collateral alpha-decay chains'' \cite{1949Mei01}. Thorium was bombarded with 150~MeV deuterons from the Berkeley 184-inch cyclotron. The $\alpha$-decay chain from $^{226}$Pa was measured following chemical separation. ``Although the mass type has not yet been identified through known daughters as above, general considerations with regard to the method of formation and half-life of the parent substance, and the energies of all the members of the series suggest a collateral branch of the 4n$+$2 family: $_{91}$Pa$^{226}\overset{\alpha}{\rightarrow}_{89}$Ac$^{222}\overset{\alpha}{\rightarrow}_{87}$Fr$^{218}\overset{\alpha}{\rightarrow}_{85}$At$^{214}\overset{\alpha}{\rightarrow}_{88}$Bi$^{210}$(RaE).''
The currently accepted half-lives for the ground state and an isomeric state of $^{222}$Ac are 5.0(5)~s and 63(3)~s, respectively.

\subsection*{$^{223,224}$Ac}
In ``Artificial collateral chains to the thorium and actinium families,'' Ghiorso et al.\ discovered $^{223}$Ac and $^{224}$Ac in 1948 \cite{1948Ghi01}. Thorium targets were irradiated with 80~MeV deuterons from the Berkeley 184-inch cyclotron. The $\alpha$-decay chains beginning at $^{227}$Pa and $^{228}$Pa were measured following chemical separation. ``Prominent soon after bombardment are a number of alpha-particle groups, which decay with the 38-minute half-life of the protactinium parent. These are due to the following collateral branch of the 4n+3 radioactive family: $_{91}$Pa$^{227}\stackrel{\alpha}{\longrightarrow}_{89}$Ac$^{223}\stackrel{\alpha}{\longrightarrow}_{87}$Fr$^{219}\stackrel{\alpha}{\longrightarrow}_{85}$At$^{215}\stackrel{\alpha}{\longrightarrow}$... After the decay of the above-described series, a second group of alpha-particle emitters can be resolved. This second series, which decays with the 22-hour half-life of its protactinium parent, is a collateral branch of the 4n radioactive family as follows: $_{91}$Pa$^{228}\stackrel{\alpha}{\longrightarrow}_{89}$Ac$^{224}\stackrel{\alpha}{\longrightarrow}_{87}$Fr$^{220}\stackrel{\alpha}{\longrightarrow}_{85}$At$^{216}\stackrel{\alpha}{\longrightarrow}$...'' The decay energies and half-lives of the decay chains were listed in a table, assigning half-lives of $\sim$2~min and $\sim$2.5~h to $^{223}$Ac and $^{224}$Ac, respectively. The currently adopted half-lives for $^{223}$Ac and $^{224}$Ac are 2.10(5)~min and 2.78(17)~h, respectively.

\subsection*{$^{225}$Ac}
Hagemann et al.\ discovered $^{225}$Ac in 1947 in ``The (4n+1) radioactive series: the decay products of U$^{233}$'' \cite{1947Hag01}. The half-lives and $\alpha$- and $\beta$-decay energies of the nuclides in the decay chain of $^{233}$U were measured. ``These decay products, which constitute a substantial fraction of the entire missing, 4n+1, radioactive series are listed together with their radioactive properties, in [the table].'' The measured half-life of 10~d agrees with the presently accepted value of 10.0(1)~d. Hagemann et al.\ acknowledge the simultaneous observation of $^{221}$Fr by English et al.\ which was submitted only a day later and published in the same issue of Physical Review on the next page \cite{1947Eng01}. A previous assignment of a 42~h half-life to $^{225}$Ac \cite{1936Ron01} was evidently incorrect.

\subsection*{$^{226}$Ac}
In the 1950 article ``Extension of alpha- and beta-decay systematics of protactinium isotopes,'' Meinke and Seaborg reported the discovery of $^{226}$Ac \cite{1950Mei01}. A thick thorium target was bombarded with 60~MeV deuterons from the Berkeley 184-in.\ cyclotron. $^{226}$Ac was identified by measuring $\alpha$ particles from the $^{226}$Th daughter following chemical separation. ``The actinium fraction from the protactinium containing 1.3$\times$10$^6$ beta-disintegrations/min.\ of Pa$^{230}$ contained 300 alpha-disintegrations/min.\ of Th$^{226}$ which decayed with the 29-hr.\ half-life of its beta-emitting Ac$^{226}$ parent.'' The quoted half-life agrees with the currently adopted value of 29.37(12)~h. Meinke and Seaborg did not consider their measurement a new discovery referring to a private communication with K.\ Street, Jr.\ and unpublished data as referenced in the 1948 Table of Isotopes \cite{1948Sea01}.

\subsection*{$^{227}$Ac}
Giesel reported the observation of a new active substance later identified as $^{227}$Ac in the 1902 paper ``Ueber Radium und radioactive Stoffe'' \cite{1902Gie01}. A raw sample consisting of barium, strontium, calcium and small amount of radium was chemical separated and radiation was observed on a zinc sulfide screen. ``Es soll nunmehr nach dem die Emanation veranlassenden K\"orper gesucht werden, der besonderer chemischer Natur zu sein scheint, da kein anderer der bekannten, stark activen Stoffe oder Thorerde eine Emanation in dieser Weise verr\"ath. Der K\"orper findet sich nach Abtrennung seltener Erden aus Pechblende mit Oxals\"aure in der folgenden Ammoniakf\"allung.'' [This emanation-producing body will be sought further; it seems to have unusual chemical properties, for neither the thorium earths nor any of the other known active materials give off this kind of emanation. The body is found in the ammonia precipitate which follows the oxalic acid separation of the rare earths from pitchblende.] (Translation from reference \cite{1971Kir01}). Two years later Giesel named the new element emanium \cite{1904Gie01}. Emanium turned out to be the same substance that Debierne had earlier called actinium \cite{1899Deb01} and Debierne is generally credited with the discovery of actinium. However, Kirby argued that Debierne could not have observed actinium in his first papers \cite{1899Deb01,1900Deb01} and thus credit should be given to Giesel \cite{1971Kir01}. The presently adopted half-life is 21.772(3)~y.

\subsection*{$^{228}$Ac}
In 1908 Hahn reported the first observation of mesothorium 2 later identified as $^{228}$Ac in the article ``Ein kurzlebiges Zwischenprodukt zwischen Mesothor und Radiothor'' \cite{1908Hah01}. Alpha- and beta-ray decay curves from a mesothorium ($^{228}$Ra) source were measured following chemical separation. ``Im Laufe der weiteren Untersuchungen hat sich nun gezeigt, da\ss\ das Mesothorium keine einheitliche Substanz ist, sondern aus zwei, in genetischem Zusammenhange stehenden Bestandteilen sich zusammensetzt; und zwar zerf\"allt das langlebige eigentliche Mesothorium ohne erkennbare Strahlenabgabe in ein neues Produkt, das die oben erw\"ahnten $\beta$-Strahlen aussendet und eine Zerfallsperiode von 6,20 Stunden besitzt.'' [It was shown in further studies that mesothorium is not a single substance, but that it consists of two, genetically connected substances; the original long-lived mesothorium decays without any visible radiation into a new product, which emits the above mentioned $\beta$-rays with a half-life of 6.20 hours.] This half-life agrees with the presently adopted value of 6.15(2)~h.

\subsection*{$^{229}$Ac}
In ``Preparation of Ra$^{229}$ and Ac$^{229}$'' Depocas and Harvey reported the discovery of $^{229}$Ac in 1952 \cite{1952Dep01}. $^{228}$Ra was irradiated with neutrons in the Chalk River NRX pile. Beta-decay curves were measured with an end-window Geiger tube following chemical separation. ``Periods of 66$\pm$5~minutes and about 6.1~hours, attributed to Ac$^{229}$ and Ac$^{228}$, respectively, were observed.'' This value agrees with the currently accepted half-life of 62.7(5)~min. Previous half-life assignments of 11~min \cite{1935Hah01}, 10$-$12~min \cite{1935Cur01,1935Cur03}, 18~min, 20$-$30~h \cite{1938Mei01}, and 3.5~h \cite{1935Cur03,1938Mei01} to $^{229}$Ac were evidently incorrect.

\subsection*{$^{230,231}$Ac}
Chayawattanangkur et al.\ reported the discovery of $^{230}$Ac and $^{231}$Ac in the 1973 paper ``Heavy isotopes of actinium: $^{229}$Ac, $^{230}$Ac, $^{231}$Ac and $^{232}$Ac'' \cite{1973Cha01}. $^{232}$Th was irradiated with 150~MeV bremsstrahlung $\gamma$ rays from the Mainz electron linear accelerator. Beta- and gamma-ray spectra were measured with a stilbene crystal and a Ge(Li) diode, respectively, following chemical separation. ``For $^{230}$Ac a half-life of 80$\pm$10~sec and $\gamma$-rays of 454.5(100) and 508.1~keV (58) were obtained. $^{231}$Ac was found to decay with a half-life of 7.5$\pm$0.1~min.'' The measured half-life for $^{230}$Ac is close to the currently adopted value of 122(3)~s and the $^{231}$Ac half-life corresponds to the presently accepted values. The previous assignment of a 15(1)~min half-life to $^{231}$Ac \cite{1960Tak01} was probably not right because the measured $\gamma$-ray intensities were incorrect \cite{2008ENS01}.

\subsection*{$^{232}$Ac}
Gippert et al.\ identified $^{232}$Ac in the 1986 article ``Decay studies of neutron-rich radium and actinium isotopes, including the new nuclides $^{232}$Ra and $^{232,234}$Ac'' \cite{1986Gip01}. An 11.4~MeV/u $^{238}$U beam from the GSI UNILAC bombarded $^{nat}$W/Ta targets producing $^{232}$Ac in multinucleon-transfer reactions. Reaction products were mass separated and $\beta$-delayed $\gamma$-ray spectra were measured with two Ge detectors. ``The half-life of $^{232}$Ac was determined from the $\beta$-, $\gamma$- and X-rays to be 119(5)~s.'' The quoted value is the currently adopted half-life. A previously measured 35(5)~s half-life \cite{1973Cha01} was evidently incorrect.

\subsection*{$^{233}$Ac}
In 1983 Chu and Zhou identified $^{233}$Ac in ``Identification of $^{233}$Ac'' \cite{1983Chu01}. Uranium targets were bombarded with 28~GeV protons from the Brookhaven AGS. Gamma-ray spectra were measured with a Ge(Li) detector following chemical separation. ``The half-life derived for $^{233}$Ac from several experiments is 2.3$\pm$0.3~min.'' This value agrees with the currently accepted half-life of 145(10)~s.

\subsection*{$^{234}$Ac}
Gippert et al.\ identified $^{234}$Ac in the 1986 article ``Decay studies of neutron-rich radium and actinium isotopes, including the new nuclides $^{232}$Ra and $^{232,234}$Ac'' \cite{1986Gip01}. An 11.4~MeV/u $^{238}$U beam from the GSI UNILAC bombarded $^{nat}$W/Ta targets producing $^{234}$Ac in multinucleon-transfer reactions. Reaction products were mass separated and $\beta$-delayed $\gamma$-ray spectra were measured with two Ge detectors. ``The $\gamma$-spectra of the A=234 sources show new $\gamma$ lines of 44(7)~s half-life, besides known $\gamma$-rays from $^{234m,g}$Pa.'' This half-life is the currently adopted value.

\subsection*{$^{235}$Ac}
The first observation of $^{235}$Ac was reported in 2006 by Bosch et al.\ in the paper ``Experiments with stored exotic nuclei at relativistic energies'' \cite{2006Bos01}. A relativistic $^{238}$U beam from the GSI SIS was fragmented and identified with the FRS. The fragments were injected into the ESR for Schottky Mass Spectrometry and lifetime measurements. The data for $^{235}$Ac were displayed in a figure: ``Discovery of the new isotope $^{235}$Ac along with its mass and lifetime measurements applying time-resolved SMS. The mass value has been extracted by calibrating with the known mass for $^{235}$Th, whereas the half-life has been extracted from the time evolution of the peak-area.'' The measured half-life of 60(4)~s is the currently adopted half-life.

\subsection*{$^{236}$Ac}
In the 2010 paper ``Discovery and investigation of heavy neutron-rich isotopes with time-resolved Schottky spectrometry in the element range from thallium to actinium'', Chen et al.\ described the discovery of $^{236}$Ac \cite{2010Che01}. A beryllium target was bombarded with a 670~MeV/u $^{238}$U beam from the GSI heavy-ion synchrotron SIS and projectile fragments were separated with the fragment separator FRS. The mass and half-life of $^{236}$Ac was measured with time-resolved Schottky Mass Spectrometry in the storage-cooler ring ESR. ``In this experiment the new isotopes of $^{236}$Ac, $^{224}$At, $^{221}$Po, $^{222}$Po, and $^{213}$Tl were discovered.'' The half-life of 72$^{+345}_{-33}$~s was listed in a table and is currently the only measured value for $^{236}$Ac.


\section{$^{208-238}$Th}\vspace{0.0cm}

The element thorium was discovered by Berzelius in 1829 \cite{1829Ber01}. The radioactivity of thorium was observed for the first time in 1898 by Schmidt \cite{1898Sch01} followed by M. Curie \cite{1898Cur01} only three months later. A detailed timeline of the discovery of thorium radioactivity can be found in the 1966 article by Badash \cite{1966Bad01}.

\begin{figure}
	\centering
	\includegraphics[scale=.7]{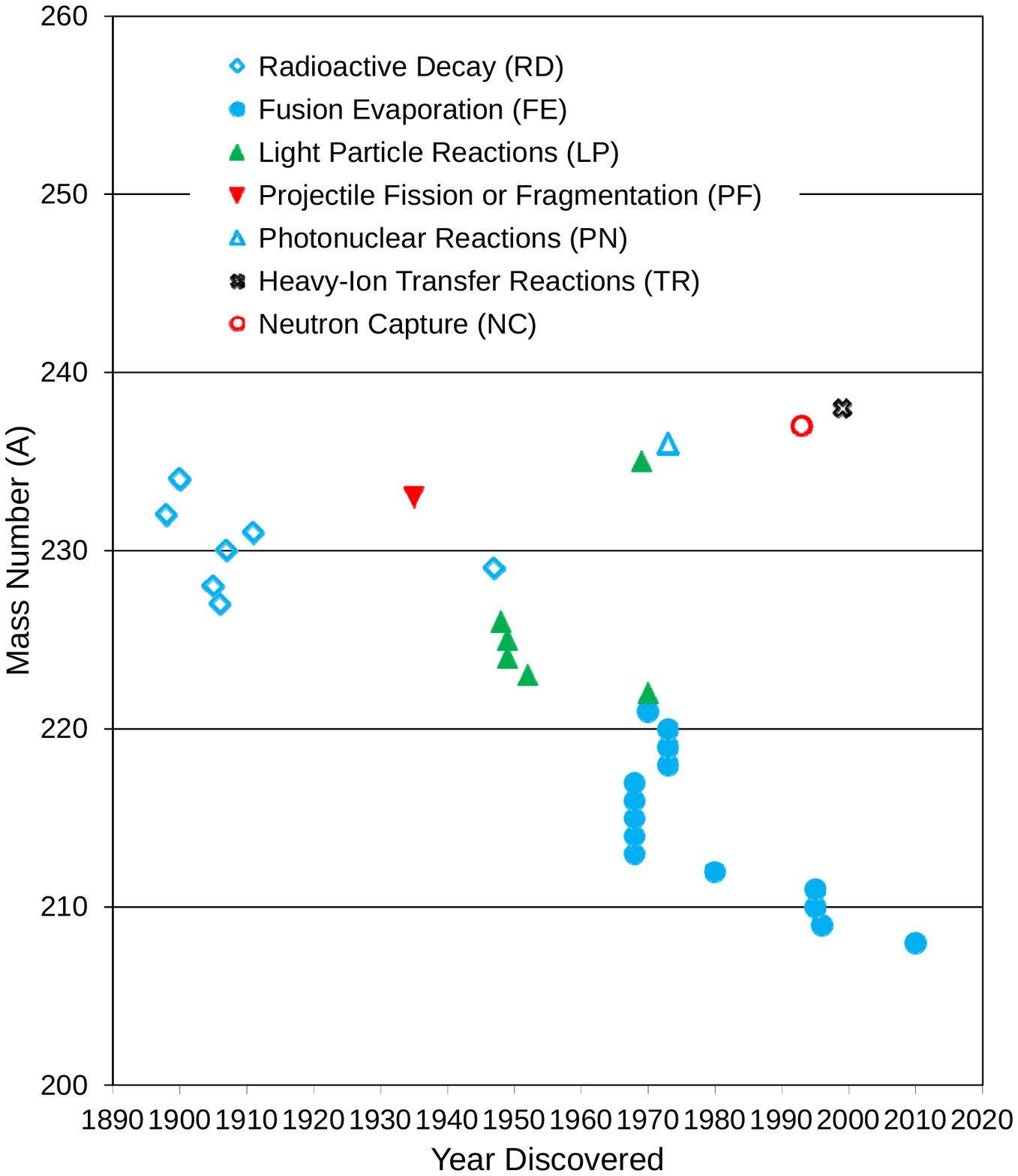}
	\caption{Thorium isotopes as a function of time when they were discovered. The different production methods are indicated.}
\label{f:year-thorium}
\end{figure}

Thirty-one thorium isotopes from A = 206--236 have been discovered so far and according to the HFB-14 model \cite{2007Gor01} about 70 additional thorium isotopes could exist. Figure \ref{f:year-thorium} summarizes the year of first discovery for all thorium isotopes identified by the method of discovery: radioactive decay (RD), fusion evaporation reactions (FE), light-particle reactions (LP), projectile fragmentation (PF), neutron capture (NC), photo-nuclear (PN), and heavy-ion transfer reactions (TR). In the following, the discovery of each thorium isotope is discussed in detail and a summary is presented in Table 1.

\subsection*{$^{208}$Th}
Heredia et al.\ identified $^{208}$Th in the 2010 paper ``The new isotope $^{208}$Th'' \cite{2010Her01}. $^{64}$Ni beams at energies of 288 and 294~MeV from the GSI UNILAC bombarded an enriched $^{147}$SmF$_{3}$ target forming $^{208}$Th in (3n) fusion-evaporation reactions. Reaction products were separated with the velocity filter SHIP and implanted in a silicon detector which also recorded subsequent $\alpha$-decay. ``Four decay chains, shown in [the table], were found. The decay properties of the last three members of these chains are in good agreement with the published values of the descendants of $^{208}$Th, i.e. the isotopes $^{204}$Ra, $^{200}$Rn and $^{196}$Po. An $\alpha$-decay energy of 8044(30)~keV and a half-life of 1.7$^{+1.7}_{-0.6}$~ms were deduced for $^{208}$Th.'' This half-life corresponds to the currently adopted value. In an earlier measurement only an upper limit for the production cross section of $^{208}$Th was given \cite{2006Kur01}.


\subsection*{$^{209}$Th}
The discovery of $^{209}$Th was reported by Ikezoe et al.\ in the 1996 paper ``$\alpha$ decay of a new isotope $^{209}$Th'' \cite{1996Ike01}. A $^{182}$W target was bombarded with a 171~MeV $^{32}$S beam from the JAERI-tandem accelerator producing $^{209}$Th in (5n) fusion-evaporation reactions. Residues were separated in flight with the JAERI-RMS recoil mass separator and implanted into a position sensitive strip detector which also recorded subsequent $\alpha$ decay. ``We conclude that the observed two decay chains listed in [the table] correspond to the $\alpha$ decays from new isotope $^{209}$Th.'' The measured half-life of 3.8$^{+69}_{-15}$~ms was recently improved to 2.5$^{+17}_{-7}$~ms \cite{2010Her01}.

\subsection*{$^{210,211}$Th}
In 1995, Uusitalo et al.\ described the observation of $^{210}$Th and $^{211}$Th in ``$\alpha$ decay of the new isotopes $^{210}$Th and $^{211}$Th'' \cite{1995Uus01}. $^{35}$Cl beams with energies of 5.2$-$5.7~MeV/nucleon from the Jyv\"askyl\"a K-130 cyclotron bombarded a $^{181}$Ta target to form $^{210}$Th and $^{211}$Th in (6n) and (5n) fusion-evaporation reactions, respectively. Recoil products were separated with the gas-filled recoil separator RITU and implanted in a position sensitive PIPS detector which also measured subsequent $\alpha$ decay. ``The measured $\alpha$ energies of $^{211}$Th and $^{210}$Th are (7792$\pm$14) and (7899$\pm$17)~keV, respectively. The half-lives were found to be (37$^{+28}_{-11}$)~ms ($^{211}$Th) and (9$^{+17}_{-4}$)~ms ($^{210}$Th).'' The $^{210}$Th half-life was recently improved to 16.0(36)~ms \cite{2010Her01} and the half-life for $^{211}$Th is included in the calculation of the currently adopted half-life.

\subsection*{$^{212}$Th}
In the 1980 paper ``$^{212}$Th, a new isotope,'' Vermeulen et al.\ identified $^{212}$Th \cite{1980Ver01}. A $^{176}$Hf target was bombarded with a 179~MeV $^{40}$Ar beam from the GSI heavy-ion accelerator UNILAC forming $^{212}$Th in (4n) fusion-evaporation reactions. Recoil products were separated with the velocity filter SHIP and implanted in a silicon surface barrier detector which also measured subsequent $\alpha$ decay. ``The half-life of $^{212}$Th was deduced from the spectrum of time distances between the impinging evaporation residues and their $\alpha$-decays using the maximum likelihood method and found to be t$_{1/2}$=(30$^{+20}_{-10}$)~ms.'' This half-life was recently improved to 31.7(13)~ms \cite{2010Her01}.

\subsection*{$^{213-217}$Th}
$^{213}$Th, $^{214}$Th, $^{215}$Th, $^{216}$Th, and $^{217}$Th were discovered in 1968 by Valli and Hyde in ``New isotopes of thorium studied with an improved helium-jet recoil transport apparatus'' \cite{1968Val02}. $^{16}$O beams with a maximum energy of 166~MeV from the Berkeley heavy ion linear accelerator HILAC bombarded $^{206}$Pb targets to produce $^{213-217}$Th in (9-5n) fusion-evaporation reactions. Recoil products were deposited on a metallic surface in front of a semiconductor detector with a helium gas jet. ``Thorium-216: ...In [the figure] an $\alpha$ group is presented at 7.921$\pm$0.008~MeV. The measured decay period for this group is 28$\pm$2~msec. Supporting evidence for the assignment of this activity to $^{216}$Th comes from the excitation function shown in [the figures] which has the shape characteristic of a compound nucleus reaction product... Thorium-215: ...We assign three groups in our spectra to $^{215}$Th: 7.522$\pm$0.008~MeV, (40$\pm$3)\%; 7.393$\pm$0.008~MeV, (52$\pm$3)\%; and 7.331$\pm$0.010~MeV, (8$\pm$3)\%. A half-life of 1.2$\pm$0.2~sec was measured for all three groups...  Thorium-214 and Thorium 213: ...At a beam energy of 142~MeV an $\alpha$ energy of 7.680$\pm$0.010~MeV and a half-life of 125$\pm$25~msec were found and assigned to $^{214}$Th. At a beam energy of 157~MeV an $\alpha$ energy of 7.690$\pm$0.010~MeV and a half-life of 150$\pm$ 25~msec were obtained and assigned to $^{213}$Th... Thorium-217: ...in half-life measurements that were run continuously during and between beam bursts, a very short-lived group was observed at 9.250$\pm$0.010~MeV... We obtained an upper limit of 0.3~msec for the half-life... The yield maximum at 120-MeV beam energy corresponds to an excitation energy of 57~MeV for the compound nucleus $^{222}$Th. This is a proper amount for the evaporation of five neutrons in the de-excitation process, indicating strongly that the 9.250-MeV activity belongs to $^{217}$Th.'' The $^{215}$Th half-life of 1.2(2)~s is the currently accepted value and the half-lives for $^{213}$Th (150(25)~ms), $^{214}$Th (125(25)~ms), and $^{216}$Th (28(2)~ms) are included in the calculation of the currently adopted values of 144(21)~ms, 87(1)~ms, and 26.0(2)~ms, for these isotopes, respectively. The upper limit of 0.3~ms for $^{217}$Th is consistent with the present value of 240(5)~$\mu$s.

\subsection*{$^{218}$Th}
Hiruta et al.\ reported the discovery of $^{218}$Th in the 1973 paper ``Alpha-particle decay of $^{218}$Th, a new isotope'' \cite{1973Hir01}. A $^{209}$Bi target was bombarded with 65$-$96~MeV $^{14}$N beams from the RIKEN cyclotron forming $^{218}$Th in the (5n) fusion-evaporation reaction. Alpha-particle spectra were measured with a Si surface barrier detector. ``Moreover, the excitation function of the 7.13~MeV peak, which is assigned to the decay of $^{214}$Ra from its energy and half-life is almost identical with that of the 9.67~MeV peak, indicating that this nucleus is populated from the decay of $^{218}$Th.'' The measured half-life of 96(7)~ns is included in the weighted average of the currently adopted value. Less than a month later H\"ausser et al.\ independently reported a half-life of 122(8)~ns for $^{218}$Th \cite{1973Hau01}.

\subsection*{$^{219,220}$Th}
The 1973 paper ``Short-lived $\alpha$ emitters of thorium: new isotopes $^{218-220}$Th'' described the discovery of $^{219}$Th and $^{220}$Th by H\"ausser et al.\ \cite{1973Hau01}. $^{16}$O beams from the Chalk River MP tandem accelerator bombarded $^{206-208}$Pb targets. $^{219}$Th was formed in the reactions $^{206}$Pb($^{16}$O,3n) and $^{207}$Pb($^{16}$O,4n) and $^{220}$Th was formed in the reactions $^{207}$Pb($^{16}$O,3n) and $^{208}$Pb($^{16}$O,4n). Reaction products were stopped in a carbon catcher foil and $\alpha$ particles were observed in an annular surface barrier detector. ``[The figure] shows a time spectrum for the 9.34-MeV $\alpha$ group from $^{219}$Th corresponding to a half-life T$_{1/2}$=1.05$\pm$0.03~$\mu$sec for this isotope.'' The half-life of 9.7(6)~$\mu$s for $^{220}$Th was listed in a table. These half-lives are the currently accepted values.

\subsection*{$^{221,222}$Th}
In 1970, Torgerson and Macfarlane reported the first observation of $^{221}$Th and $^{222}$Th in ``Alpha decay of the $^{221}$Th and $^{222}$Th decay chains'' \cite{1970Tor01}. A 10.6~MeV/nucleon $^{16}$O beam from the Yale heavy ion accelerator was used to bombard a $^{208}$Pb target forming $^{221}$Th and  $^{222}$Th in (3n) and (2n) fusion-evaporation reactions, respectively. Recoil products were transported to a stainless steel surface with a helium jet and $\alpha$ spectra were measured with a Si(Au) surface barrier detector. ``On the basis of excitation function and half-life data, the groups observed at 8.472, 8.146 and 7.733~MeV are assigned to $^{221}$Th. Decay curves for these activities have been measured and the half-life of $^{221}$Th was found to be 1.68$\pm$0.06~msec... A weak 7.984~MeV $\alpha$-particle transition having a half-life of 4$\pm$1~msec has been assigned to $^{222}$Th.'' The half-life of 1.68(6)~ms corresponds to the presently accepted value for $^{221}$Th and the $^{222}$Th half-life of 4(1)~ms agrees with the currently adopted value of 2.8(3)~ms. Only three days later Valli et al.\ independently reported half-lives of 1.8(3)~ms and 2.8(3)~ms for $^{221}$Th and $^{222}$Th, respectively.

\subsection*{$^{223}$Th}
In 1952, $^{223}$Th was discovered by Meinke et al.\ and the results were reported in the paper ``Further work on heavy collateral radioactive chains'' \cite{1952Mei01}. Thorium nitrate targets were irradiated with a $^4$He beam from the Berkeley 184-inch cyclotron. $^{227}$U was chemically separated and the energy of $\alpha$-particles were measured with an alpha-particle pulse analyzer. ``An additional short-lived chain collateral to the actinium (4n+3) natural radioactive family has also been partially identified. This chain decays as follows: U$^{227}\rightarrow$Th$^{223}\rightarrow$Ra$^{219}\rightarrow$Em$^{215}\rightarrow$Po$^{211}\rightarrow$Pb$^{207}$.''
The currently accepted half-life of $^{223}$Th is 600(20)~ms.

\subsection*{$^{224,225}$Th}
Meinke et al.\ reported the observation of $^{224}$Th and $^{225}$Th in the 1949 paper ``Three additional collateral alpha-decay chains'' \cite{1949Mei01}. Thorium was bombarded with 100$-$120~MeV $^4$He beams from the Berkeley 184-inch cyclotron. Alpha-decay chains from $^{228}$U and $^{229}$U were measured following chemical separation. ``The irradiation of thorium with 100-Mev helium ions resulted in the observation of the following collateral branch of the artificial 4n$+$1, neptunium, radioactive family shown with Po$^{213}$ and its decay products: $_{92}$U$^{229}\overset{\alpha}{\rightarrow}_{90}$Th$^{225}\overset{\alpha}{\rightarrow}_{88}$Ra$^{221}\overset{\alpha}{\rightarrow}_{86}$Em$^{217}\ldots$ Immediately after 120-Mev helium ion bombardment of thorium the uranium fraction contains another series of five alpha-emitters, which is apparently a collateral branch of the 4n family: $_{92}$U$^{228}\overset{\alpha}{\rightarrow}_{90}$Th$^{224}\overset{\alpha}{\rightarrow}_{88}$Ra$^{220}\overset{\alpha}{\rightarrow}_{86}$Em$^{216}\ldots$'' In a table summarizing the energies and half-lives of the decay chain only the $\alpha$-decay energy was given for $^{224}$Th stating a calculated half-life of $\sim$1~s. The currently accepted half-life is 1.05(2)~s. The measured half-life of 7.8(3)~min for $^{225}$Th agrees with the presently adopted value of 8.72(4)~min.

\subsection*{$^{226}$Th}
Studier and Hyde reported the discovery of $^{226}$Th in the 1948 paper ``A new radioactive series - the protactinium series'' \cite{1948Stu01}. Thorium metal targets were bombarded with 19~MeV deuterons and a 38~MeV $^4$He beam from the Berkeley 60-inch cyclotron forming $^{230}$Pa in (d,4n) and ($\alpha$,p5n) reactions. $^{226}$Th was populated by subsequent $\alpha$ decay after the initial $\beta^-$ decay of $^{230}$Pa to $^{230}$U. Alpha-decay spectra were measured following chemical separation. ``The half-life obtained for Th$^{226}$ is 30.9~min.'' The quoted half-life is close to the currently adopted value of 30.57(10)~min.

\subsection*{$^{227}$Th}
A new radioactive substance later identified as $^{227}$Th was identified by Hahn in the 1906 paper ``A new product of actinium'' \cite{1906Hah01}. Alpha- and beta-ray activities of an actinium solution were measured following chemical separation. ``...I have found that a new product is present in actinium which is intermediate between actinium and actinium X, and, from analogy to thorium, will be called for convenience `radio-actinium.' This product emits $\alpha$ rays, is half-transformed in about twenty days, and is the parent of actinium X.'' The currently adopted half-life is 18.68(9)~d.

\subsection*{$^{228}$Th}
In 1905 Hahn published ``A new radio-active element, which evolves thorium emanation. Preliminary communication'' reporting the observation of new radioactive substance later identified as $^{228}$Th \cite{1905Hah01}. Activities from a ``thorianite'' sample were measured following chemical separation. ``By a series of troublesome operations, a quantity of precipitate was obtained by aid of ammonia, and to separate iron, it was treated in acid solution with ammonium oxalate; this produced about 10 milligrammes of crystalline precipitate, which was by far the most active preparation obtained, and which shows after two months no diminution in its radio-active power... The close relation of the new body to thorium is proved, not merely by the apparent identity of the two emanations, but also in its having been separated from a mineral unusually rich in thorium.'' The currently accepted half-life is 1.9116(16)~y.

\subsection*{$^{229}$Th}
Hagemann et al.\ discovered $^{229}$Th in 1947 in ``The (4n+1) radioactive series: the decay products of U$^{233}$'' \cite{1947Hag01}. The half-lives and $\alpha$- and $\beta$-decay energies of the nuclides in the decay chain of $^{233}$U were measured. ``These decay products, which constitute a substantial fraction of the entire missing, 4n+1, radioactive series are listed together with their radioactive properties, in [the table].'' The measured half-life of 7$\times$10$^3$~y agrees with the presently accepted value of 7880(120)~y. Hagemann et al.\ acknowledge the simultaneous observation of $^{229}$Th by English et al.\ which was submitted only a day later and published in the same issue of Physical Review on the next page \cite{1947Eng01}.

\subsection*{$^{230}$Th}
``The origin of radium'' reported the 1907 discovery of a new radioactive substance later identified $^{230}$Th by Boltwood \cite{1907Bol01}. Thorium was extracted from uranium minerals and $\alpha$-activities were measured following chemical separation. ``For these and certain other reasons I think that there is good cause for believing that uranium minerals contain an element emitting $\alpha$ rays, which is different from the other elements that have been identified, which produces no emanation, and which resembles thorium in its chemical properties.'' Later in the year Boltwood suggested the name ``ionium'' for the new substance \cite{1907Bol02,1907Bol03}. The presently accepted half-life of $^{230}$Th is 75380(300)~y.


\subsection*{$^{231}$Th}
The 1911 discovery of UrY later identified as $^{231}$Th was reported in ``The disintegration products of uranium'' by Antonoff \cite{1911Ant01}. Alpha- and beta-ray activities were measured from a uranium nitrate sample following chemical separation. ``The period of the new product deduced from the curve is 1.5~days. A number of other curves obtained in a similar manner gave about the same period of decay. It is proposed to call the new product uranium Y (UrY).'' The currently adopted half-life of $^{231}$Th is 25.52(1)~h.

\subsection*{$^{232}$Th}
Schmidt reported the radioactivity thorium later identified as $^{232}$Th in the 1898 paper ``Ueber die von den Thorverbindungen und einigen anderen Substanzen ausgehende Strahlung'' \cite{1898Sch01}. The activity was observed with photographic plates and an electroscope. ``Von den vielen Elementen und Verbindungen, welche ich hieraufhin gepr\"uft habe, hat sich nur eines gefunden, welches sich dem Uran analog verh\"alt, n\"amlich das Thor... Es scheint als ob dieselben an das hohe Atomgewicht Uran = 240, Thorium = 232 gebunden sind.'' [Out of the many elements and compounds which I have subsequently tested, only one - thorium - behaved analogous to uranium... It seems that these are related to the high atomic weight, 240 for uranium and 232 for thorium.] Less than three months later \cite{1966Bad01} M. Curie independently reported the radioactivitiy of thorium \cite{1898Cur01}. The atomic weight of thorium was not shown to be close to mass 232 until 1909 \cite{1909Cla02} and Aston demonstrated in 1932 that $^{232}$Th constituted the only thorium isotope of primordial origin \cite{1932Ast02}. The presently adopted half-life is 1.405(6)$\times$10$^{10}$~y.

\subsection*{$^{233}$Th}
In 1935, the discovery of $^{233}$Th was reported by Hahn and Meitner in the paper ``Die k\"unstliche Umwandlung des Thoriums durch Neutronen: Bilding der bisher fehlenden radioaktiven 4n + 1-Reihe'' \cite{1935Hah01}. Thorium was irradiated with neutrons with and without paraffin. ``Von dieser Substanz konnte durch chemisehe Trennungen nachgewiesen werden, da\ss\ sie mit dem Thorium isotop ist. Sie stellt also offenbar das durch Einfangen eines Neutrons entstandene Thorisotop Th$^{233}$ dar, und ihre Bildung wird daher, wie oben gezeigt, durch langsame Neutronen erheblich beg\"unstigt.'' [By chemical separation it could be demonstrated that this substance is isotopic with thorium. It thus represents the thorium isotope $^{233}$Th produced by neutron capture and its formation could be enhanced by slow neutrons as shown above.] This half-life agrees with the presently adopted value of 22.3(1)~min. A 15~min \cite{1934Fer01} and a 24~min \cite{1935Ama01} were reported earlier without mass assignment. Less than a month later I. Curie et al.\ assigned the 25~min half-life to $^{233}$Th without giving credit to Hahn and Meitner although quoting the paper \cite{1935Cur01}.

\subsection*{$^{234}$Th}
Crookes reported the discovery of a radioactive substance later identified as $^{234}$Th in the 1900 paper ``Radio-activity of uranium'' \cite{1900Cro01}. Activities of uranium nitrate were detected following chemical separation. ``Having thus definitely proved that the supposed radio-activity of uranium and its salts is not an inherent property of the element, but is due to the presence of a foreign body, it is necessary patiently to determine the nature of the foreign body [For the sake of lucidity the new body must have a name. Until it is more tractable I will call it provisionally UrX$-$the unknown substance of uranium]. Several radio-active bodies claimed to be new have already been extracted from pitchblende, and experiments have been instituted to see if the newly found body UrX had similar chemical properties to those of older substances.'' UrX was later renamed UrX$_1$. The currently accepted half-life is 24.10(3)~d.

\subsection*{$^{235}$Th}
In the 1969 paper ``Wirkungsquerschnitte der (n,p)-, (n,pn)-, und (n,$\alpha$)-Reaktionen am Uran-238 mit 15-Mev-Neutronen und Nachweis des Thoriums-235,'' Trautmann et al.\ reported the discovery of $^{235}$Th \cite{1969Tra01}. Uranyl nitrate was irradiated with 14.8~MeV T-D-neutrons produced with the Mainz cascade accelerator forming $^{235}$Th in the reaction $^{238}$U(n,$\alpha$). Beta-decay curves were measured following chemical separation. ``A new isotope of thorium, $^{235}$Th with a half-life of 6.9$\pm$0.2~min, was found by repeated milking of $^{235}$Pa.'' This half-life agrees with the presently accepted value of 7.2(1)~min. An earlier attempt to measure the half-life of $^{235}$Th was unsuccessful \cite{1950Har01}.

\subsection*{$^{236}$Th}
The first observation of $^{236}$Th was reported by Kaffrell and Trautmann in ``Identification of $^{236}$Th'' in 1973 \cite{1973Kaf01}. $^{238}$U targets were irradiated with bremsstrahlung $\gamma$-rays with a maximum energy of 140~MeV produced by bombarding a thick copper target with electrons from the Mainz linear accelerator. Gamma-ray spectra and decay curves were measured following chemical separation. ``For the mass assignment and determination of the $^{236}$Th half-life, the genetic relationship to the well-known 9.1~min $^{236}$Pa has been used. This facilitated the differentiation between $^{236}$Th and the lighter thorium nuclides formed under the described irradiation conditions by ($\gamma$,2pxn) reactions having higher cross sections. An analysis of the growth-decay curve of the strongest $\gamma$-ray of $^{236}$Pa at 642~keV yields a half-life of 36$\pm$3~min for $^{236}$Th.'' This value is included in the calculation of the currently adopted half-life of 37.3(15)~min. Seven months later Orth et al.\ independently reported a half-life of 37.5(15)~min \cite{1973Ort02}.

\subsection*{$^{237}$Th}
Yuan et al.\ identified $^{237}$Th in the 1993 paper ``The synthesis and identification of new heavy neutron-rich nuclide $^{237}$Th'' \cite{1993Yua01}. $^{238}$U was irradiated with 14~MeV neutrons produced by bombarding a TiT target with deuterons from the Lanzhou 600-kV Cockcroft-Walton accelerator. Gamma-ray spectra and decay curves were measured with a GMX HPGe detector following chemical separation. ``A radioactive-series decay analyzing program was applied resulting in the half-lives of 5.0$\pm$0.9~min and 8.5$\pm$1.0~min for $^{237}$Th and $^{237}$Pa, respectively.'' This value is included in the calculation of the currently adopted half-life of 4.8(5)~min.

\subsection*{$^{238}$Th}
In 1999, He et al.\ reported the observation of $^{238}$Th in ``Synthesis and identification of a new heavy neutron-rich isotope $^{238}$Th'' \cite{1999He01}. A natural uranium target was bombarded with a 60~MeV/u $^{18}$O beam and $^{238}$Th was produced in a multi-nucleon transfer reactions from $^{238}$U. X- and $\gamma$-ray spectra and decay curves were measured with two HPGe detectors following chemical separation. ``Half-lives of 9.4$\pm$2.0~min for $^{238}$Th and 2.1$\pm$0.4~min for $^{238}$Pa were extracted from the data of [the figure] using a computer code for analyzing the decay of a radioactive series.'' This half-life of 9.4(20)~min is the currently accepted value for $^{238}$Th.


\section{$^{212-239}$Pa}\vspace{0.0cm}

Protactinium was discovered in 1913 by Fajans and G\"ohring with the observation of $^{234}$Pa \cite{1913Faj01}. Fajans and G\"ohring originally named the new substance UrX$_2$, however, in a later article Fajans and Beer chose the name Brevium: ``Die Veranlassung, dem kurzlebigen, von O. G\"ohring und mir entdeckten Element einen besonderen Namen zu geben, liegt darin, da\ss\ dieses Element das einzige ist, dem die bis jetzt freie Stelle in der f\"unften Gruppe der letzten Horizontalreihe des periodischen Systems zukommt. F\"ur seine radioaktive Charakterisierung wird die Bezeichnung UrX$_2$ vorzuziehen sein, weil sie sofort dessen genetische Beziehung klarlegt.'' [The reason to name the short-lived element which O. G\"ohring and I discovered is that it is the only element which fits in the location of the fifth group of the last horizontal row of the period table. However, for the radioactive characterization the name UrX$_2$ is preferred, because it immediately clarifies its genetic relation.] \cite{1913Faj02}.

In 1918, Soddy and Cranston \cite{1918Sod01} and Hahn and Meitner \cite{1918Hah01} independently discovered $^{231}$Pa. They are generally given equal credit for the discovery \cite{1933Wee01}. Hahn and Meitner named the new substance protactinium \cite{1918Hah01,1918Mei01}, however, Soddy quoted Hahn and Meitner with the name protoactinium \cite{1918Sod03,1918Sod02}. Both names were subsequently in use until the name protactinium was officially selected at the 15$^{th}$ IUPAC conference in Amsterdam in 1949 \cite{1949IUP01,2005Kop01}.

Once it was demonstrated that brevium and protactinium belonged to the same element, Fajans withdraw the right to name the element: ``Dem Protactinium kommt im periodischen System dieselbe Stelle zu, wie dem Uran X$_2$, und da es als das bei weitem langlebigere Element den Namen der Plejade bestimmt, besteht f\"ur die Bezeichnung Brevium kein Bed\"urfnis mehr.'' [Protactinium occurs in the periodic system in the same location as uranium X$_2$, and since it is the by far longer-lived element which determines the name, the term breviums no longer needed.] \cite{1920Faj01}. In 1973, Fajans and Morris reflected on the discovery and naming of protactinium \cite{1973Faj01}.

\begin{figure}
	\centering
	\includegraphics[scale=.7]{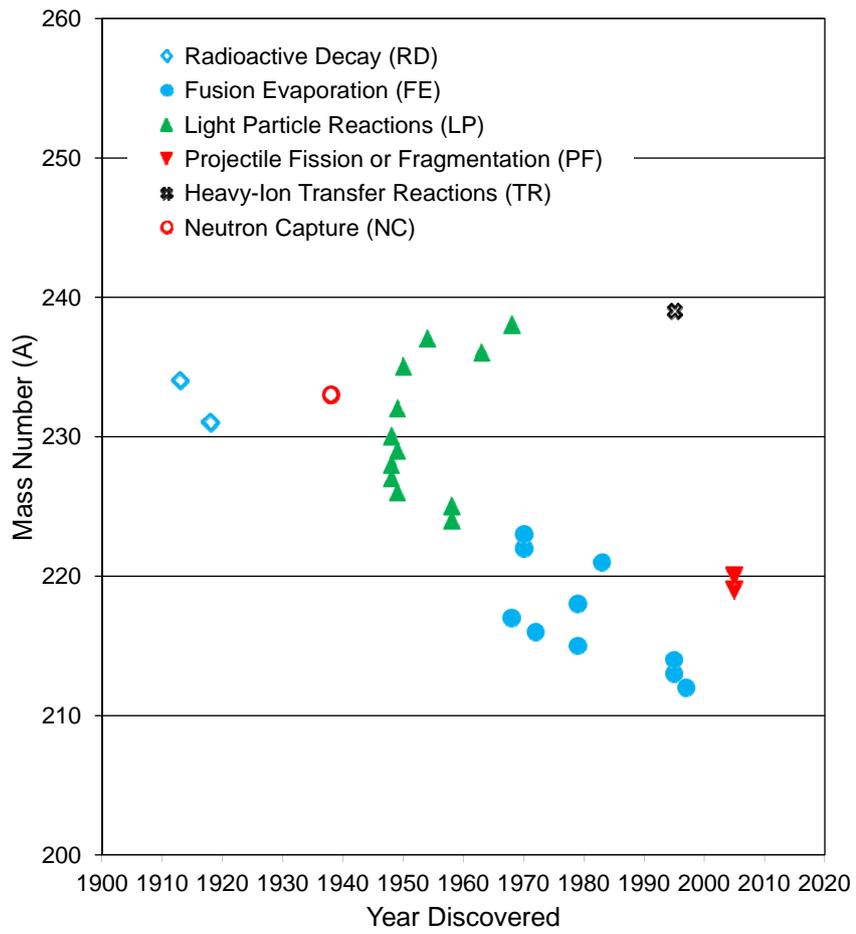}
	\caption{Protactinium isotopes as a function of time when they were discovered. The different production methods are indicated.}
\label{f:year-protactinium}
\end{figure}

Twenty-eight protactinium isotopes from A = 212--239 have been discovered so far and according to the HFB-14 model \cite{2007Gor01} about 60 additional protactinium isotopes could exist. Figure \ref{f:year-protactinium} summarizes the year of first discovery for all protactinium isotopes identified by the method of discovery: radioactive decay (RD), fusion evaporation reactions (FE), light-particle reactions (LP), projectile fragmentation (PF), neutron capture (NC), and heavy-ion transfer reactions (TR). In the following, the discovery of each protactinium isotope is discussed in detail and a summary is presented in Table 1.

\subsection*{$^{212}$Pa}
In the 1997 paper ``$\alpha$-decay properties of the new neutron deficient isotope $^{212}$Pa'' Mitsuoka et al.\ identified $^{212}$Pa \cite{1997Mit01}. A 182.5~MeV $^{35}$Cl beam from the JAERI-tandem accelerator bombarded a target of $^{182}$W forming $^{212}$Pa in the (5n) fusion-evaporation reaction. Reaction products were separated in flight with the JAERI recoil mass separator and implanted into a double-sided position sensitive silicon detector which also measured subsequent $\alpha$-decay. ``The $\alpha$ decay from the ground state of $^{212}$Pa has been observed with an $\alpha$-particle energy of 8.270(30)~MeV and a half-life of 5.1$^{+6.1}_{-1.9}$~ms.'' This value is the currently accepted half-life.

\subsection*{$^{213,214}$Pa}
The discovery of $^{213}$Pa and $^{214}$Pa was reported by Ninov et al.\ in the 1995 paper ``Identification of the neutron-deficient isotopes $^{213,214}$Pa'' \cite{1995Nin01}. The GSI UNILAC accelerator was used to bombard enriched $^{170}$Er targets with 5.2$-$5.6~AMeV $^{51}$V beams. Evaporation residues were separated in-flight with the velocity filter SHIP and implanted in a position sensitive PIPS detector which also recorded subsequent $\alpha$ decay. ``From the time differences between implantation of the evaporation residues and the subsequent $\alpha$-decays a half-life of T$_{1/2}$=(5.3$^{+4.0}_{-1.6}$)~ms was measured for $^{213}$Pa, while a value of T$_{1/2}$=(17$\pm$3)~ms was obtained for $^{214}$Pa.'' The quoted values for both isotopes are the currently adopted half-lives.

\subsection*{$^{215}$Pa}
In the 1979 article ``Alpha decay properties of new protactinium isotopes,'' Schmidt et al.\ described the observation of $^{215}$Pa \cite{1979Sch01}. A tantalum target was bombarded with 165$-$202~MeV $^{40}$Ar beams from the GSI heavy-ion accelerator UNILAC and $^{215}$Pa was formed in the fusion-evaporation reaction $^{181}$Ta($^{40}$Ar,6n). Residues were separated with the velocity filter SHIP and implanted in a counter telescope consisting of a transmission secondary electron detector and a silicon surface barrier detector. ``$^{215}$Pa was found to decay with E$_\alpha$ = 8.085$\pm$0.015~MeV and t$_{1/2}$=14$\pm^{20}_{3}$~ms.'' This half-life agrees with the currently adopted value of 14(2)~ms.

\subsection*{$^{216}$Pa}
Sung-Ching-Yang et al.\ reported the observation of $^{216}$Pa in the 1972 paper ``Production of the new isotope Pa$^{216}$ by bombardment of osmium by phosphorus and of gold by magnesium'' \cite{1972Sun01}. Enriched $^{189}$Os and $^{190}$Os targets were bombarded with $\sim$200~MeV $^{31}$P beams and $^{187}$Au targets were bombarded with 146~MeV $^{24}$Mg beams from the Dubna 310-cm heavy-ion cyclotron. $^{216}$Pa was formed in the reactions $^{189}$Os($^{31}$P,4n), $^{190}$Os($^{31}$P,5n), and $^{197}$Au($^{24}$Mg,5n). Recoils were slowed in a gas chamber and deposited in front of two Si(Au) detectors. ``In the same figure we have shown the decay curve of the $\alpha$ emitter with E$_\alpha$ equal to 7.72, 7.82, and 7.92~MeV. It is evident that the exponential corresponds to a half-life T$_{1/2}$=0.20$\pm$0.04~sec.'' This value agrees with the currently adopted half-life of 0.15$^{+6}_{-4}$~s.

\subsection*{$^{217}$Pa}
The observation of $^{217}$Pa was first reported in 1968 by Valli and Hyde in ``New isotopes of thorium studied with an improved helium-jet recoil transport apparatus'' \cite{1968Val02}. $^{20}$Ne beams with a maximum energy of 166~MeV from the Berkeley heavy ion linear accelerator HILAC bombarded $^{203}$Tl and $^{206}$Pb targets to produce $^{217}$Pa in (6n) and (1p8n) fusion-evaporation reactions, respectively. Recoil products were deposited on a metallic surface in front of a semiconductor detector with a helium gas jet. ``We have observed a weak short-lived $\alpha$ group at 8.340$\pm$0.010~MeV in spectra obtained by bombardment of $^{203}$Tl and $^{208}$Pb with $^{20}$Ne ions, which on the basis of the evidence, can be assigned to $^{217}$Pa.'' The currently adopted half-life is 3.48(9)~ms.

\subsection*{$^{218}$Pa}
In the 1979 article ``Alpha decay properties of new protactinium isotopes,'' Schmidt et al.\ described the observation of $^{218}$Pa \cite{1979Sch01}. A tantalum target was bombarded with 165$-$202~MeV $^{40}$Ar beams from the GSI heavy-ion accelerator UNILAC and $^{218}$Pa was formed in the fusion-evaporation reaction $^{181}$Ta($^{40}$Ar,3n). Residues were separated with the velocity filter SHIP and implanted in a counter telescope consisting of a transmission secondary electron detector and a silicon surface barrier detector. ``For $^{218}$Pa two $\alpha$-lines were found at 9.535$\pm$0.015~MeV and 9.614$\pm$0.020~MeV with a corresponding half-life of 120$^{+40}_{-20}$~$\mu$s.'' This half-life agrees with the currently adopted value of 113(10)~$\mu$s.

\subsection*{$^{219,220}$Pa}
Liu et al.\ reported the observation of $^{219}$Pa and $^{220}$Pa in the 2005 paper ``Decay spectroscopy of suburanium isotopes following projectile fragmentation of $^{238}$U at 1 GeV/u'' \cite{2005Liu01}. A 1~GeV/u $^{238}$U beam from the GSI heavy-ion synchrotron SIS bombarded a beryllium target forming $^{219}$Pa and$^{220}$Pa in projectile fragmentation reactions. Reaction products were separated with the FRS fragment separator and implanted into a Si telescope which measured subsequent $\alpha$ decay. The production yields for protactinium, thorium, and actinium isotopes were shown in a figure: ``The dips at $^{219}$Pa(T$_{1/2}$ = 53~ns), $^{218}$Th(T$_{1/2}$ = 109~ns) and $^{217}$Ac(T$_{1/2}$ = 69~ns) are due to the fact that their half-lives are shorter than their time of flight from target to S4.'' These discoveries were not specifically mentioned, but the observation of $^{219}$Pa and $^{220}$Pa were shown in a `yield per day' plot. In the figure caption a half-life of $^{220}$Pa is listed referring to the 1996 Table of Isotopes \cite{1996Fir01} which cites only a 1987 conference proceeding \cite{1987Fae01}. For $^{220}$Pa the ENSDF database refers to the 1987 conference proceeding \cite{1987Fae01} and an internal report \cite{1987Miy01}.
The presently accepted half-lives for $^{219}$Pa and $^{220}$Pa are 53(10)~ns and 780(160)~ns, respectively.

\subsection*{$^{221}$Pa}
$^{221}$Pa was first observed by Hingmann et al.\ and the results were published in the 1983 paper ``Identification of $^{222}$U and $^{221}$Pa by $\alpha$-correlation chains'' \cite{1983Hin01}. A $^{186}$W target was bombarded with a 4.72~MeV/u $^{40}$Ar beam from the GSI linear accelerator UNILAC to form $^{221}$Pa in (p4n) fusion-evaporation reactions. Evaporation residues were separated by the velocity filter SHIP and implanted in a surface barrier detector which measured subsequent $\alpha$ decays. ``Altogether six $\alpha$-cascades could be assigned to $^{221}$Pa, and the evaluation of the measured decay times yields a half-life of (6.1$^{+3.7}_{-2.4}$)~$\mu$s for this isotope.'' The given value agrees with the currently adopted half-life of 5.9(17)~$\mu$s.

\subsection*{$^{222,223}$Pa}
In the 1970 article ``Production and decay properties of protactinium isotopes of mass 222 to 225 formed in heavy-ion reactions,'' Borggreen et al.\ identified $^{222}$Pa and $^{223}$Pa \cite{1970Bor01}. The Berkeley heavy-ion linear accelerator (HILAC) was used to bombard $^{209}$Bi, $^{208}$Pb and $^{205}$Tl targets with $^{16}$O, $^{19}$F and $^{22}$Ne beams forming $^{224}$Pa and $^{223}$Pa in (xn) fusion-evaporation reactions. Recoil products were deposited by a helium gas stream on a metal surface located in front of a gold surface-barrier detector which recorded the subsequent $\alpha$ decay. ``From the combined results a value of 6.5$\pm$1.0~msec was determined for the $^{223}$Pa half-life... The 8.18-MeV $^{222}$Pa peak and the 9.21-MeV $^{218}$Ac peak provided the best data leading to a value of 5.7$\pm$0.5~msec for the $^{222}$Pa half-life.'' The half-life for $^{222}$Pa of 5.7(5)~ms is within a factor of two of the present value of 2.9$^{+0.6}_{-0.4}$~s and the half-life for $^{223}$Pa of 6.5(10)~ms is included in the calculation of the currently accepted value of 5.1(6)~ms.

\subsection*{$^{224,225}$Pa}
The first detection of $^{224}$Pa and $^{225}$Pa was reported in 1958 by Tove in ``Alpha-emitters with short half-life induced by protons on heavy elements'' \cite{1958Tov01}. Thorium targets were bombarded with 170~MeV protons from the Uppsala synchrocyclotron forming $^{224}$Pa and $^{225}$Pa in the reactions $^{232}$Th(p,9n) and $^{232}$Th(p,8n), respectively. Alpha-particle spectra were measured with a 180$^\circ$ single-focusing magnetic spectrometer and decay curves were recorded. ``From the plot of half-life the most probable half-lives for the different series are 1.05$\pm$0.05~sec. for Th$^{224}$, 0.6$\pm$0.05~sec. for Pa$^{224}$, and for Pa$^{225}$ and Th$^{223}$ 0.8$\pm$0.1 and 0.9$\pm$0.1~sec. respectively.'' The half-life of 0.60(5)~s for $^{224}$Pa agrees with the presently adopted value of 0.844(19)~s and the half-life of 0.8(1)~s for $^{225}$Pa is within a factor of two of the accepted value of 1.7(2)~s.

\subsection*{$^{226}$Pa}
Meinke et al.\ reported the observation of $^{226}$Pa in the 1949 paper ``Three additional collateral alpha-decay chains'' \cite{1949Mei01}. Thorium was bombarded with 150~MeV deuterons from the Berkeley 184-inch cyclotron. The $\alpha$-decay chain from $^{226}$Pa was measured following chemical separation. ``Similarly the protactinium fraction of 150-MeV deuteron-bombarded thorium shows a series of alpha-particle emitters whose rate of decay is controlled by the 1.7-minute half-life of the parent with the subsequent members all too short-lived to be isolated and separately studied. Although the mass type has not yet been identified through known daughters as above, general considerations with regard to the method of formation and half-life of the parent substance, and the energies of all the members of the series suggest a collateral branch of the 4n$+$2 family: $_{91}$Pa$^{226}\overset{\alpha}{\rightarrow}_{89}$Ac$^{222}\overset{\alpha}{\rightarrow}_{87}$Fr$^{218}\overset{\alpha}{\rightarrow}_{85}$At$^{214}\overset{\alpha}{\rightarrow}_{88}$Bi$^{210}$(RaE).'' The measured half-life of 1.70(15)~min for $^{226}$Pa agrees with the currently accepted value of 1.8(2)~min.

\subsection*{$^{227,228}$Pa}
In ``Artificial collateral chains to the thorium and actinium families,'' Ghiorso et al.\ discovered $^{227}$Pa and $^{228}$Pa in 1948 \cite{1948Ghi01}. Thorium targets were irradiated with 80~MeV deuterons from the Berkeley 184-inch cyclotron. The $\alpha$-decay chains beginning at $^{227}$Pa and $^{228}$Pa were measured following chemical separation. ``Prominent soon after bombardment are a number of alpha-particle groups, which decay with the 38-minute half-life of the protactinium parent. These are due to the following collateral branch of the 4n+3 radioactive family: $_{91}$Pa$^{227}\stackrel{\alpha}{\longrightarrow}_{89}$Ac$^{223}\stackrel{\alpha}{\longrightarrow}_{87}$Fr$^{219}\stackrel{\alpha}{\longrightarrow}_{85}$At$^{215}\stackrel{\alpha}{\longrightarrow}$... After the decay of the above-described series, a second group of alpha-particle emitters can be resolved. This second series, which decays with the 22-hour half-life of its protactinium parent, is a collateral branch of the 4n radioactive family as follows: $_{91}$Pa$^{228}\stackrel{\alpha}{\longrightarrow}_{89}$Ac$^{224}\stackrel{\alpha}{\longrightarrow}_{87}$Fr$^{220}\stackrel{\alpha}{\longrightarrow}_{85}$At$^{216}\stackrel{\alpha}{\longrightarrow}$...'' The half-lives of 38~min and 22~h for $^{227}$Pa and $^{228}$Pa agree with the currently adopted values of 38.3(3)~min and 22(1)~h, respectively.

\subsection*{$^{229}$Pa}
The discovery of $^{229}$Pa by Hyde et al.\ was described in the 1949 paper ``A new isotope of protactinium: Pa$^{229}$'' \cite{1949Hyd02}. A thorium target was bombarded with 22~MeV deuterons from the Berkeley 60-in.\ cyclotron forming $^{229}$Pa in the reaction $^{230}$Th(d,3n). After chemical separation, $\alpha$-ray spectra were measured with a pulse-analyzer. ``A search in the protactinium fraction for unidentified activities that might be due to Pa$^{229}$ and Pa$^{228}$ revealed the presence of a previously unknown $\alpha$ emitter. The decay of this activity could not be followed directly because U$^{230}$ and its daughters were growing into the protactinium fraction. However, it was possible to follow the decay of the unknown activity by making frequent measurements of the $\alpha$-ray spectrum of the protactinium fraction with a pulse analyzer and by following the decay of the unknown peak. From such measurements a half life of 1.4$\pm$0.4~days was determined.'' This value agrees with the currently adopted half-life of 1.50(5)~d.

\subsection*{$^{230}$Pa}
Studier and Hyde reported the discovery of $^{230}$Pa in the 1948 paper ``A new radioactive series - the protactinium series'' \cite{1948Stu01}. Thorium metal targets were bombarded with 19~MeV deuterons and a 38~MeV $^4$He beam from the Berkeley 60-inch cyclotron forming $^{230}$Pa in (d,4n) and ($\alpha$,p5n) reactions. Alpha-decay spectra were measured following chemical separation. ``A preliminary value was obtained by extracting and measuring the U$^{230}$ produced in equal periods of time from a sample of protactinium. In this manner a crude half-life value of about two weeks was obtained. A more accurate value was obtained by mounting a sample of Pa$^{230}$ on a platinum counting disk and following the rate of change of the alpha-activity of the sample... In this way using three separate samples a value of 17.0$\pm$0.5~days was determined for the half-life of Pa$^{230}$.'' This half-life is included in the calculation of the currently accepted value of 17.4(5)~d.

\subsection*{$^{231}$Pa}
In 1918, Hahn and Meitner reported the discovery of protactinium ($^{231}$Pa) in ``Die Muttersubstanz des Actiniums, ein neues radioaktives Element von langer Lebensdauer'' \cite{1918Hah01}. The $\alpha$-activity of a uranium sample was measured following chemical separation. ``Zusammenfassung der Resultate: 1. Die bisher hypothetische Muttersubstanz des Actiniums wurde aufgefunden und in radioaktiv reinem Zustande an Erds\"auren konzentriert hergestellt. Sie ist ein h\"oheres Homologes des Tantals. 2. Sie sendet $\alpha$-Strahlen vom Durchdringungsbereich 3,14~cm aus. 3. Die Halbwertszeit betr\"agt mindestens 1,200 und h\"ochstens 180,000 Jahre... 6. F\"ur das neue radioaktive Element wurde der Name Protactinium gew\"ahlt.'' [Summary of the results: 1. The previously hypothetical parent of actinium was discovered and produced in a radioactively pure state concentrated in transition metals. It is a higher homologue of tantalum. 2. It emits $\alpha$-rays with a penetration depth of 3.14~cm. 3. The half-life is at least 1,200 and at most 180,000 years... 6. The name protactinium was chosen for the new radioactive element.] This half-life estimate is consistent with the presently accepted value of 32,760(110)~y. Less than three months earlier, Soddy and Cranston had reported indirect evidence for the existence of $^{231}$Pa \cite{1918Sod01}.

\subsection*{$^{232}$Pa}
The first observation of $^{232}$Pa was published in 1949 by Gofman and Seaborg in ``Production and properties of U$^{232}$ and Pa$^{232}$ \cite{1949Gof01}. A $^{232}$Th target was bombarded with 14~MeV deuterons from the Berkeley 60-in.\ cyclotron producing $^{232}$Pa in (d,2n) reactions. Beta- and gamma-rays were measured with a Lauritsen quartz-fiber electroscope and alpha-particles were measured in an ionization chamber following chemical separation. ``It was found that the deuteron bombardment of thorium produces a 1.6-day $\beta$- and $\gamma$-emitting $_{91}$Pa$^{232}$, which decays to a 30-year $\alpha$-emitting $_{92}$U$^{232}$.'' This value for $^{232}$Pa agrees with the currently accepted half-life of 1.32(2)~d.

\subsection*{$^{233}$Pa}
Meitner et al.\ reported the discovery of $^{233}$Pa in the 1938 paper ``K\"unstliche Umwandlungsprozesse bei Bestrahlung des Thoriums mit Neutronen; Auftreten isomerer Reihen durch Abspaltung von $\alpha$-Strahlen'' \cite{1938Mei01}. A thorium sample surrounded by paraffin was irradiated with neutrons from a Ra-Be source. Decay curves were measured following chemical separation. ``[Die Figur] zeigt die Abklingkurve der gereinigten Protactiniumpr\"aparate. Die Halbwertszeiten liegen zwischen 24 und 27 Tagen.'' [[The figure] shows the decay curves of the separated protactinium samples. The half-lives are between 24 and 27 days.] This half-life agrees with the presently adopted value of 26.967(2)~d. A previously reported half-life of 2.5~min \cite{1935Cur02} was evidently incorrect.

\subsection*{$^{234}$Pa}
In the 1913 article ``\"Uber die komplexe Natur des UrX,'' Fajans and G\"ohring identified UX$_2$ ($^{234}$Pa) \cite{1913Faj01}. Beta-rays from a UX ($^{234}$Th) solution were measured following chemical separation. ``Die Versuche, die wir zur Pr\"ufung dieses Schlusses angestellt haben, f\"uhrten uns in der Tat zur Auffindung eines neuen Elementes (UrX$_2$) mit einer Halbwertzeit von ca. 1.1 Min., dem, wie es scheint, die harten $\beta$-Strahlen des UrX zukommen.'' [The experiments which we performed to test this conclusion, indeed lead to the discovery of a new element (UrX$_2$) with a halflife of approximately 1.1~min. The hard $\beta$-rays of UrX seem to originate from this new element.] The measured half-life agrees with the presently adopted value of 1.17(3)~min for an isomeric state. The ground state of $^{234}$Pa was discovered by Hahn in 1921 who named it at first Z \cite{1921Hah01} and then UZ \cite{1921Hah02} and its half-life is 6.70(5)~h.

\subsection*{$^{235}$Pa}
In the 1950 article ``Extension of alpha- and beta-decay systematics of protactinium isotopes,'' Meinke and Seaborg reported the discovery of $^{235}$Pa \cite{1950Mei01}. $^{238}$U was bombarded with 19~MeV deuterons and 9.5~MeV protons from the Berkeley 60-in.\ cyclotron. Beta-decay curves were measured following chemical separation. ``These observations are consistent with the assignment of the 23.7-min.\ activity to Pa$^{235}$, produced in the reactions U$^{238}$(d,$\alpha$n)Pa$^{235}$ and U$^{238}$(p,$\alpha$)Pa$^{235}$.'' This value agrees with the currently adopted half-life of 24.44(11)~min.

\subsection*{$^{236}$Pa}
Wolzak and Morinaga reported the observation of $^{236}$Pa in the 1963 article ``Protactinium-236'' \cite{1963Wol01}. Uranyl nitrate was bombarded with 26~MeV deuterons from the Amsterdam cyclotron forming $^{236}$Pa in the reaction $^{238}$U(d,$\alpha$). Beta- and gamma-rays were measured with an anthracene crystal and a beryllium window NaI crystal, respectively, following chemical separation. ``On the basis of the assumption that this component is due to the decay of $^{236}$Pa, the $^{238}$U(d,$\alpha$)$^{236}$Pa-reaction cross-section was calculated to be roughly of the order of 100 $\mu$b. This value is considered to be reasonable for this reaction. We are therefore inclined to assign the 12-minute activity to $^{236}$Pa.'' This measured 12.5(10)~min half-life is close to the currently adopted half-life of 9.1(1)~min.

\subsection*{$^{237}$Pa}
In the 1954 paper ``New isotope protactinium-237,'' Crane and Iddings described the observation of $^{237}$Pa \cite{1954Cra01}. A $^{238}$U target was bombarded with 190~MeV deuterons from the Berkeley 184-in.\ cyclotron forming $^{237}$Pa in the (d,2pn) reaction. Beta-ray decay curves were measured following chemical separation. ``From this series of experiments, one obtains a half-life of 10.5$\pm$1~min for Pa$^{237}$.'' This half-life agrees with the currently adopted value of 8.7(2)~min.

\subsection*{$^{238}$Pa}
Trautmann et al.\ published the discovery of $^{238}$Pa in ``Heavy isotopes of protactinium'' in 1968 \cite{1968Tra01}.  Natural uranium targets were irradiated with 14.8~MeV T-D-neutrons produced with the Mainz Cockcroft-Walton accelerator forming $^{238}$Pa in the charge-exchange reaction $^{238}$U(n,p). Following chemical separation, $\beta$- and $\gamma$-ray spectra were measured with a plastic scintillator and a Ge(Li) detector, respectively. ``The $\gamma$-ray spectra of 2.3-min $^{238}$Pa is shown in [the figure]. Eighty-four $\gamma$-transitions are observed with energies and intensities as indicated in the figure.'' The measured half-life of 2.3(1)~min is included in the weighted average to determine the currently adopted value of 2.27(9)~min.

\subsection*{$^{239}$Pa}
In 1995, Yuan et al.\ reported the observation of $^{239}$Pa in ``A new isotope of protactinium: $^{239}$Pa'' \cite{1995Yua01}.  A natural uranium target was bombarded with a 50~MeV/u $^{18}$O beam and $^{239}$Pa was produced in a multi-nucleon transfer reactions. X- and $\gamma$-ray spectra and decay curves were measured with two HPGe detectors following chemical separation. ``A radioactive-series decay analyzing program was applied and half-lives of 106$\pm$30~min and 22$\pm$5~min for $^{239}$Pa and $^{239}$U, respectively were extracted.'' This value is the currently adopted half-life.


\section{$^{217-242}$U}\vspace{0.0cm}

The element uranium was discovered by Klaproth in 1789 \cite{1789Kla01} and the radioactivity of uranium was observed for the first time in 1896 by Becquerel \cite{1896Bec01}.

\begin{figure}
	\centering
	\includegraphics[scale=.7]{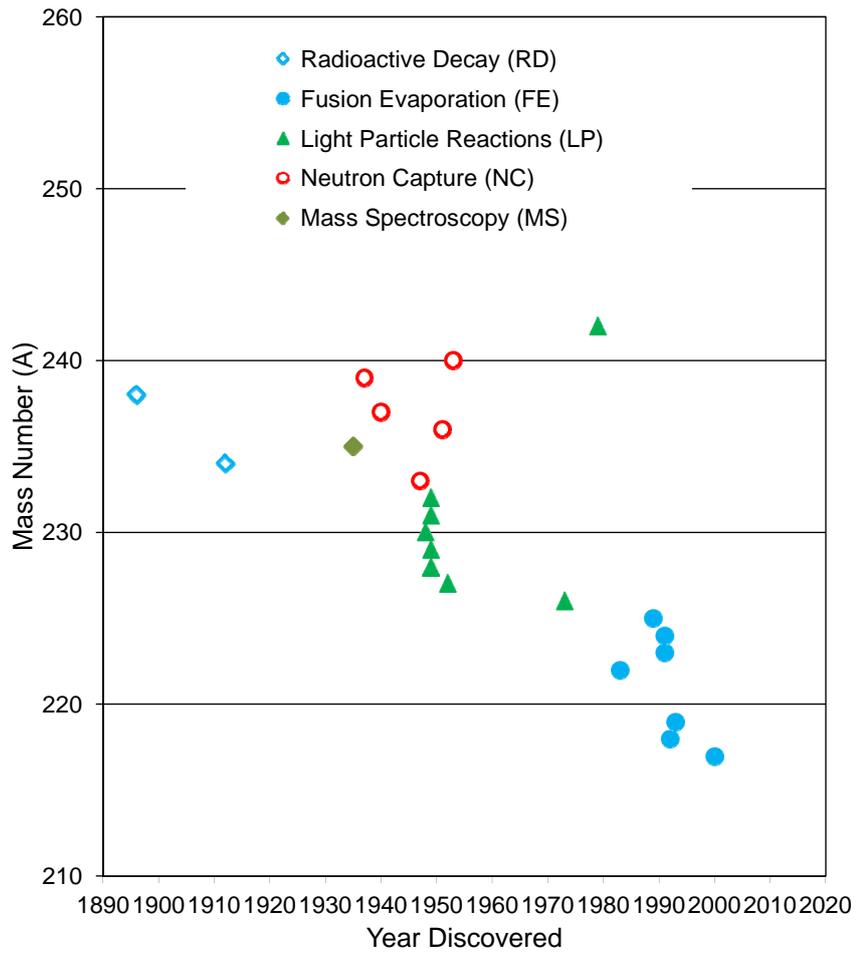}
	\caption{Uranium isotopes as a function of time when they were discovered. The different production methods are indicated.}
\label{f:year-uranium}
\end{figure}

Twenty-three uranium isotopes from A = 217--242 have been discovered so far with $^{220,221}$U and $^{241}$U yet to be observed. According to the HFB-14 model \cite{2007Gor01} about 80 additional uranium isotopes could exist. Figure \ref{f:year-uranium} summarizes the year of first discovery for all uranium isotopes identified by the method of discovery: radioactive decay (RD), fusion evaporation reactions (FE), light-particle reactions (LP), mass spectroscopy (MS), and neutron capture reactions (NC). In the following, the discovery of each uranium isotope is discussed in detail and a summary is presented in Table 1.

\subsection*{$^{217}$U}
Malyshev et al.\ reported the first observation of $^{217}$U in the 2000 paper ``The new isotope $^{217}$U'' \cite{2000Mal01}. A 193~MeV $^{40}$Ar beam from the Dubna U-400 cyclotron bombarded a $^{182}$W target forming $^{217}$U in the (5n) fusion-evaporation reaction. Residues were separated in flight with the VASSILISSA recoil separator and implanted in a silicon position-sensitive strip-detector array which also recorded subsequent $\alpha$ decay. ``A group of three events with a mean energy of 8005$\pm$20~keV was attributed to the decay of the new isotope $^{217}$U. These events are listed in [the table]. The half-life of the new isotope was calculated from the time intervals of subsequent ER-$\alpha$ events and was determined to be 15.6$^{+21.3}_{-5.7}$~ms.'' This half-life was included in the calculation of the currently adopted value of 16$^{+21}_{-6}$~ms.

\subsection*{$^{218}$U}
In 1992, Andreyev et al.\ identified $^{218}$U in ``The new isotope $^{218}$U'' \cite{1992And01}. A 5.96~MeV/u $^{27}$Al beam from the Dubna U-400 cyclotron bombarded a $^{197}$Au target to produce $^{218}$U in the (6n) fusion-evaporation reaction. The residues were separated with the electrostatic separator VASSILISSA and implanted into an array of eight position sensitive silicon detectors which also recorded subsequent $\alpha$ decay. ``The $\alpha$-decay energy and the half-life of $^{218}$U were determined to be 8625$\pm$25~keV and 1.5$^{+7.3}_{-0.7}$~ms, respectively.'' This value is close to the currently adopted half-life of 0.51$^{+17}_{-10}$~ms.

\subsection*{$^{219}$U}
The discovery of $^{219}$U was reported in the 1993 article ``The new isotope $^{219}$U'' by Andreyev et al.\ \cite{1993And01}. A 5.5~MeV/u $^{27}$Al beam from the Dubna U-400 cyclotron bombarded a $^{197}$Au target forming $^{219}$U in the (5n) fusion-evaporation reaction. The residues were separated with the electrostatic separator VASSILISSA and implanted into an array of six position sensitive silicon detectors which also recorded subsequent $\alpha$ decay. ``The $\alpha$-decay energy and the half-life of $^{218}$U were determined to be 9860$\pm$40~keV and 42$^{+34}_{-13}$~ms, respectively.'' The quoted half-life is the currently adopted value.

\subsection*{$^{222}$U}
$^{222}$U was first observed by Hingmann et al.\ and the results were published in the 1983 paper ``Identification of $^{222}$U and $^{221}$Pa by $\alpha$-correlation chains'' \cite{1983Hin01}. A natural tungsten target was bombarded with a 4.5~MeV/u $^{40}$Ar beam from the GSI linear accelerator UNILAC to form $^{222}$U in (4n) fusion-evaporation reactions. Evaporation residues were separated by the velocity filter SHIP and implanted in a surface barrier detector which also measured subsequent $\alpha$ decays. ``The correlation between the two events is certain, because the rate of $^{214}$Ra $\alpha$-decays was extremely low: During the measuring time of 9~h, a total number of only twelve $^{214}$Ra $\alpha$-decays was observed. By means of the maximum likelihood method, a half-life of (1.0$^{+1.2}_{-0.4}$)~$\mu$s was deduced from the detected three $^{222}$U decays.'' This half-life is the currently accepted value.

\subsection*{$^{223,224}$U}
$^{223}$U and $^{224}$U were identified by Andreyev et al.\ in the 1991 article ``Production cross sections and decay properties of new $\alpha$ emitters: $^{223,224}$U'' \cite{1991And01}. A $^{208}$Pb target was bombarded with 101$-$125~MeV $^{20}$Ne beams from the Dubna U-400 cyclotron producing $^{223}$U and $^{224}$U in (5n) and (4n) fusion-evaporation reactions, respectively. Residues were separated with the electrostatic separator VASILISSA and implanted into a passivated ion implanted silicon detector which recorded subsequent $\alpha$ decay. ``$^{223}$U was found to decay with E$_\alpha$=(8780$\pm$40)~keV and T$_{1/2}$=18$^{+10}_{-5}$~$\mu$s. For $^{224}$U the $\alpha$-line at E$_\alpha$=(8470$\pm$15)~keV and T$_{1/2}$=0.7$^{+0.5}_{-0.2}$~ms was observed.'' The $^{223}$U half-life is the currently accepted value and the half-life of $^{224}$U is included in the weighted average to determine the currently adopted value of 0.9(3)~ms.

\subsection*{$^{225}$U}
Andreyev et al.\ reported the discovery of $^{225}$U in the 1989 paper ``Measurement of cross sections for reactions with evaporation of light particles in the complete fusion channel in bombardment of Au and Pb by Ne ions'' \cite{1989And01}. An enriched $^{208}$Pb target was bombarded with 100$-$130~MeV $^{22}$Ne beams from the Dubna cyclotron to produced $^{225}$U in (5n) fusion-evaporation reactions. The recoils were separated with the Vasilisa kinematic separator and implanted into a surface barrier detector. ``The isotope $^{225}$U was identified on the basis of correlations, well identified in the spectrum, of the transition E$^1_\alpha$=7.87~MeV and the transitions E$^2_\alpha$=8.15 and 8.47~MeV ($^{221}$Th) and 8.1~MeV ($^{213}$Rn).'' The measured half-life of 30$^{+20}_{-10}$~ms is within the uncertainties within a factor of two of the currently adopted value of 95(15)~ms. Three months later, He{\ss}berger et al.\ independently reported a half-life of 80$^{+40}_{-20}$~ms for $^{225}$U \cite{1989Hes01}.

\subsection*{$^{226}$U}
$^{226}$U was discovered by Viola et al.\ as described in the 1973 paper ``Identification of the nuclide $^{226}$U'' \cite{1973Vio01}. A $^{232}$Th target was bombarded with a 140~MeV $^4$He beam from the University of Maryland cyclotron producing $^{226}$U in (10n) reactions. Spallation products were stopped on a catcher film and  $\alpha$ decay was measured with two surface barrier semiconductor detectors. ``A least squares analysis gives a half-life of 0.50$^{+0.17}_{-0.12}$~sec. Taking into account the accuracy of the background corrections, we assign a value of 0.5$\pm$0.2 to the half-life of the 7.43~MeV activity.'' This value agrees with the currently adopted half-life of 269(6)~ms.

\subsection*{$^{227}$U}
In 1952, $^{227}$U was discovered by Meinke et al.\ and the results were reported in the paper ``Further work on heavy collateral radioactive chains'' \cite{1952Mei01}. Thorium nitrate targets were irradiated with a $^4$He beam from the Berkeley 184-inch cyclotron. $^{227}$U was chemically separated and the energy of $\alpha$-particles were measured with an alpha-particle pulse analyzer. ``An additional short-lived chain collateral to the actinium (4n+3) natural radioactive family has also been partially identified. This chain decays as follows: U$^{227}\rightarrow$Th$^{223}\rightarrow$Ra$^{219}\rightarrow$Em$^{215}\rightarrow$Po$^{211}\rightarrow$Pb$^{207}$... The half-life of the U$^{227}$ parent of the series was determined by following the decay of certain alpha-groups in the pulse analysis curves. Resolution of these decay curves into the U$^{229}$, U$^{228}$, and U$^{227}$ components gave a half-life of 1.3$\pm$0.3 minutes for U$^{227}$.'' This half-life agrees with the currently adopted value of 1.1(1)~min.

\subsection*{$^{228,229}$U}
Meinke et al.\ reported the observation of $^{228}$U and $^{229}$U in the 1949 paper ``Three additional collateral alpha-decay chains'' \cite{1949Mei01}. Thorium was bombarded with 100$-$120~MeV $^4$He beams from the Berkeley 184-inch cyclotron. Alpha-decay chains from $^{228}$U and $^{229}$U were measured following chemical separation. ``The irradiation of thorium with 100-Mev helium ions resulted in the observation of the following collateral branch of the artificial 4n$+$1, neptunium, radioactive family shown with Po$^{213}$ and its decay products: $_{92}$U$^{229}\overset{\alpha}{\rightarrow}_{90}$Th$^{225}\overset{\alpha}{\rightarrow}_{88}$Ra$^{221}\overset{\alpha}{\rightarrow}_{86}$Em$^{217}\ldots$ Immediately after 120-Mev helium ion bombardment of thorium the uranium fraction contains another series of five alpha-emitters, which is apparently a collateral branch of the 4n family: $_{92}$U$^{228}\overset{\alpha}{\rightarrow}_{90}$Th$^{224}\overset{\alpha}{\rightarrow}_{88}$Ra$^{220}\overset{\alpha}{\rightarrow}_{86}$Em$^{216}\ldots$ The 9.3-minute half-life of U$^{228}$ controls the decay rate of the series, with the half-lives of all the other members too short for them to be isolated and separately studied in our experiments.'' The measured half-life of 9.3(5)~min for $^{228}$U agrees with the presently adopted value of 9.1(2)~min. In a table summarizing the energies and half-lives of the chains, the half-life of $^{229}$U is given as 58(3)~min, which corresponds to the currently adopted value.

\subsection*{$^{230}$U}
Studier and Hyde reported the discovery of $^{230}$U in the 1948 paper ``A new radioactive series - the protactinium series'' \cite{1948Stu01}. Thorium metal targets were bombarded with 19~MeV deuterons and a 38~MeV $^4$He beam from the Berkeley 60-inch cyclotron forming $^{230}$Pa in (d,4n) and ($\alpha$,p5n) reactions. $^{230}$U was populated by subsequent $\beta$ decay. Alpha-decay spectra were measured following chemical separation. ``A U$^{230}$ decay curve is shown in [the figure]. A `least squares' treatment of the data yielded 20.8~days for the half-life.'' This half-life is the currently adopted value.

\subsection*{$^{231}$U}
In the 1949 paper ``Products of the deuteron and helium-ion bombardments of Pa$^{231}$,'' Osborne et al.\ reported the discovery of $^{231}$U \cite{1949Osb01}. A $^{231}$Pa target was bombarded with 21~MeV deuterons from the Berkeley 60-in.\ cyclotron forming $^{231}$U in (d,2n) reactions. Gamma-rays and X-rays were measured with an end-window brass-wall Geiger-Mueller counter following chemical separation. ``A new 4.2-day x-ray activity, found in the uranium fraction, has been tentatively assigned to K-electron capture by U$^{231}$.'' This value is the currently adopted half-life.

\subsection*{$^{232}$U}
The first observation of $^{232}$U was published in 1949 by Gofman and Seaborg in ``Production and properties of U$^{232}$ and Pa$^{232}$ \cite{1949Gof01}. A $^{232}$Th target was bombarded with 14~MeV deuterons from the Berkeley 60-in.\ cyclotron producing $^{232}$Pa in (d,2n) reactions. $^{232}$U was then populated by subsequent $\beta$-decay. Beta- and gamma-rays were measured with a Lauritsen quartz-fiber electroscope and alpha-particles were measured in an ionization chamber following chemical separation. ``Since U$^{232}$ is too long-lived to permit following its decay readily, its half life can best be evaluated by using the measured value of the intensity of the $\beta$ particles of the 1.6-day Pa$^{232}$, together with the corresponding measured value of the intensity of $\alpha$ particles from the daughter U$^{232}$... Since the electroscope was not calibrated for Pa$^{232}$ radiation, a rough answer could be obtained by assuming that the efficiency of the electroscope is the same for Pa$^{232}$ radiation as it is for Pa$^{233}$ radiation, for which the electroscope had been calibrated in previous work. In this manner it was found that 0.12 millicurie of Pa$^{232}$ $\beta$ radioactivity decayed to 0.016 microcurie of U$^{232}$ $\alpha$ activity. These data lead to a value of about thirty years for the half life of U$^{232}$.'' The currently accepted half-life of for $^{232}$U is 68.9(4)~y.

\subsection*{$^{233}$U}
$^{233}$U was identified by Seaborg et al.\ in the 1947 paper ``Nuclear properties of U$^{233}$: a new fissionable isotope of uranium'' \cite{1947Sea01}. A $^{232}$Th sample was irradiated with neutrons to produce $^{233}$Th in neutron capture reactions. $^{233}$U was then populated by subsequent $\beta$-decay. Fission and $\alpha$-particle decay was measured. ``We have measured the radioactive and the fission properties of U$^{233}$. Our measurements on a sample of U$^{233}$ weighing 3.8 micrograms show that this isotope undergoes fission with neutrons. The same result was obtained in a check experiment with another sample of U$^{233}$ weighing 0.8 microgram.'' The paper had been submitted in 1942: ``This paper was mailed from Berkeley, California, to the `Uranium Committee' in Washington, D.C. on April 14, 1942. The experimental work was done during 1941 and the early part of 1942. It is being published in the open literature now for historical purposes, with the original text somewhat changed, by omissions, in order to conform to present declassification standards. The information covered in this document will appear in Division IV of the MPTS, as part of a contribution of the University of California.'' In the later publication within the National Nuclear Energy Series a half-life of 1.2$\times$10$^5$~y was quoted \cite{1949Sea01} which agrees with the presently accepted value of 1.592(2)$\times$10$^5$~y.

\subsection*{$^{234}$U}
Geiger and Nuttall described the observation of a new uranium isotopes, U$_{II}$ in the 1912 paper ``The ranges of the $\alpha$ particles from uranium'' \cite{1912Gei01}. The $\alpha$-particle ranges from a uranium source were measured with a Bragg ionization chamber. ``Uranium I therefore, which has a period of 5$\times$10$^9$ years, emits $\alpha$ particles of range 2.5~cm in air at atmospheric pressure and at 15$^\circ$ C., and is followed by another $\alpha$-ray product, uranium II, which has a period of about 2$\times$10$^6$ years and emits $\alpha$ particles of range 2.9 cm.'' Earlier, the existence of a second $\alpha$ emitting uranium isotope was suspected from the number of emitted $\alpha$ particles per uranium atom \cite{1908Bol01,1910Gei01}. From the nuclear systematics of the uranium decay Soddy identified in 1923 U$_{II}$ with $^{234}$U \cite{1923Sod01}. However, it was not until 1939 that Nier showed that $^{234}$U was a very minor constituent of naturally occurring uranium with an abundance 1/17000 percent that of $^{238}$U \cite{1939Nie01}. The currently adopted half-life of $^{234}$U is 245,500(600)~y.

\subsection*{$^{235}$U}
In the 1935 article ``Isotopic constitution of uranium'' Dempster reported the discovery of $^{235}$U in natural uranium\cite{1935Dem03}. Uranium samples were used in a spark source of the Chicago mass spectrograph. ``It was found that an exposure of a few seconds was sufficient for the main component at 238 reported by Dr.\ Aston; but in in addition on long exposures a faint companion of atomic weight 235 was also present. With two different uranium electrodes it was observed on eight photographs, and two photographs with the pitchblende electrode also showed the new component.'' $^{235}$U is of primordial origin with a half-life of 7.01(1)$\times$10$^8$~y and an abundance of 0.72\%. Rutherford had predicted the existence of $^{235}$U in 1929 naming it actino-uranium \cite{1929Rut01} based on Aston's extrapolation of 231 for the mass of protactinium as the precursor of $^{207}$Pb \cite{1929Ast01}.

\subsection*{$^{236}$U}
The observation of $^{236}$U was first reported in ``The uranium isotope U$^{236}$'' by Ghiorso et al.\ in 1951 \cite{1951Ghi01}. A sample of enriched $^{235}$U was irradiated with slow neutrons. Alpha-decay was measured with an alpha-pulse analyzer apparatus following chemical separation. ``This corresponded to an alpha-half-life of U$^{236}$ of about 2$\times$10$^7$ years. Measurements, a little later, on another sample similarly prepared, containing a different concentration of U$^{235}$, led to the same result.'' This value agrees with the currently adopted half-life of 2.342$\times$10$^7$(4)~y. Four months later Jaffey et al.\ independently reported a half-life of 2.46$\times$10$^7$~y \cite{1951Jaf01}.

\subsection*{$^{237}$U}
Nishina et al.\ described the discovery of $^{237}$U in the 1940 article ``Induced $\beta$-activity of uranium by fast neutrons'' \cite{1940Nis01}. A uranium oxide sample was irradiated with fast neutrons produced by bombarding lithium with 3~MeV deuterons from the Tokyo cyclotron. Beta-decay curves were measured following chemical separation. ``The activity of the irradiated uranium was compared with that of a non irradiated sample, in order to subtract the growing $\beta$-activity due to disintegration products of uranium. The difference thus obtained shows a 6.5-day period. This activity is probably due to U$^{237}$ produced from U$^{238}$ through loss of a neutron, as in the case of the production of UY from thorium.'' This half-life agrees with the currently adopted value of 6.752(2)~d. Less than two months later McMillan independently reported a 7.0(2)~d half-life \cite{1940McM02}.

\subsection*{$^{238}$U}
In 1896 Becquerel described the discovery of radioactivity in the article ``Sur les radiations \'emises par phosphorescence'' \cite{1896Bec01}. The effect of a sample of phosphorescent uranium salt containing what was later identified as $^{238}$U on a photographic plate was studied. ``On doit donc conclure de ces exp\'eriences que la substance phosphorescente en question \'emet des radiations qui traversent le papier opaque \`a la lumi\'ere et r\'eduisent les sels d'argent.'' [From these experiments we must therefore conclude that the phosphorescent substance in question emits radiation which passes through the paper which is opaque to light and reduces the silver salts.] The atomic weight was not shown to be close to mass 238 until 1917 \cite{1917Cla01} but this allowed Soddy in 1923 \cite{1923Sod01} to identify $^{238}$U as the source of the uranium decay chain. However, only in 1931 Aston showed that $^{238}$U was the principal isotope of naturally occurring uranium \cite{1931Ast02}. $^{238}$U is of primordial origin with a half-life of 4.468(3)$\times$10$^{9}$~y.

\subsection*{$^{239}$U}
Meitner et al.\ reported the observation of $^{239}$U in the 1937 paper ``\"Uber die Imwandlungsreihen des Urans, die durch Neutronenbestrahlung erzeugt werden'' \cite{1937Mei01}. Uranium was irradiated with neutrons and $\beta$-decay curves were measured with Geiger-M\"uller counters. Three different decay paths were observed. ``Bei dem dritten Proze\ss\ haben wir nur ein Umwandlungsprodukt, ein $\beta$-strahlendes Uranisotop von 28 Min.\ Halbwertszeit nachgewiesen, aus dem unbedingt ein (vermutlich langlebiges) Ekarhenium entstehen mu\ss... Die Versuche, die Art der Ausl\"osungsprozesse festzustellen, haben ergeben, da\ss\ bei allen drei Prozessen keine $\alpha$-Abspaltung stattfindet und f\"ur alle drei Prozesse wegen der Gr\"o\ss e der gefundenen Wirkungsquerschnitte eine Herleitung vom Uran 235 und erst recht vom Uran 234 ausgeschlossen erscheint. Also m\"u\ss en die Prozesse Einfangprozesse des Uran 238 sein, was zu drei isomeren Kernen Uran 239 f\"uhrt. Dieses Ergebnis ist mit den bisherigen Kernvorstellungen sehr schwer in \"Ubereinstimmung zu bringen.'' [In the third process we found only one active product, a $\beta$-emitting uranium isotopes with a half-life of 28~min which must lead to a (probably long-lived) ekarhenium... The experiments to determine the nature of the neutron capture process demonstrated that none of the three processes exhibit $\alpha$ radiation. The magnitude of the cross sections for all three processes seem to rule out reactions on uranium 235 and even more so on uranium 234. This result is very difficult to explain within the current understanding of the nucleus.] The reported 28~min half-life agrees with the presently adopted value of 23.45(2)~min. The other decay paths most likely were due to fission products.

\subsection*{$^{240}$U}
In 1950, the discovery of $^{240}$U was described in the article ``The radiations of U$^{240}$ and Np$^{240}$'' by Knight et al.\ \cite{1953Kni01}. Uranium was irradiated with neutrons to form $^{240}$U by successive neutron capture. Following chemical separation, decay curves were measured with continuous-flow methane gas proportional counters and $\beta$- and $\gamma$-ray spectra were recorded with a magnetic lens spectrometer and a NaI(Tl) crystal, respectively. ``These measurements yielded a U$^{240}$ half-life of 14.1$\pm$0.2~hours, and a Np$^{240}$ half-life of 7.3$\pm$0.3~minutes.'' The quoted half-life agree with the currently adopted value of 14.1(1)~h. Knight et al.\ credited Hyde and Studier with the discovery of $^{240}$U quoting an unpublished report \cite{1948Hyd01}.

\subsection*{$^{242}$U}
Haustein et al.\ reported the observation of $^{242}$U in the 1979 paper ``Identification and decay of $^{242}$U and $^{242}$Np'' \cite{1979Hau01}. $^{244}$Pu targets were irradiated with 30$-$160~MeV neutrons produced at the Brookhaven Medium Energy Intense Neutron (MEIN) facility by bombarding a water-cooled copper beam stop with 200~MeV protons from the Alternating Gradient Synchrotron. Gamma- and beta-rays were measured with Ge(Li) and plastic detectors, respectively, following chemical separation. ``By combining the data from several of the most intense runs we have by least square analyses T$_{1/2}$=16.8$\pm$0.5~min for $^{242}$U and T$_{1/2}$=2.2$\pm$0.2~min for $^{242}$Np'' This half-life is the currently adopted value for $^{242}$U.


\section{Summary}
The discoveries of the known actinium, thorium, protactinium, and uranium isotopes have been compiled and the methods of their production discussed. Only the following five isotopes were initially identified incorrectly: $^{225,229,231,232}$Ac and $^{233}$Pa. In addition, the half-life of $^{233}$Th was at first reported without a definite mass assignment.

\ack

This work was supported by the National Science Foundation under grants No. PHY06-06007 (NSCL).

\bibliography{../isotope-discovery-references}

\newpage

\newpage

\TableExplanation

\bigskip
\renewcommand{\arraystretch}{1.0}

\section{Table 1.\label{tbl1te} Discovery of actinium, thorium, protactinium, and uranium isotopes }
\begin{tabular*}{0.95\textwidth}{@{}@{\extracolsep{\fill}}lp{5.5in}@{}}
\multicolumn{2}{p{0.95\textwidth}}{ }\\

Isotope & Actinium, thorium, protactinium, and uranium isotope \\
First author & First author of refereed publication \\
Journal & Journal of publication \\
Ref. & Reference \\
Method & Production method used in the discovery: \\

  & FE: fusion evaporation \\
  & NC: Neutron capture reactions \\
  & LP: light-particle reactions (including neutrons) \\
  & MS: mass spectroscopy \\
  & RD: radioactive decay \\
  & PN: photo-nuclear reactions \\
  & HI: heavy-ion transfer reactions \\
  & SP: spallation reactions \\
  & PF: projectile fragmentation \\

Laboratory & Laboratory where the experiment was performed\\
Country & Country of laboratory\\
Year & Year of discovery \\
\end{tabular*}
\label{tableI}

\datatables 



\setlength{\LTleft}{0pt}
\setlength{\LTright}{0pt}


\setlength{\tabcolsep}{0.5\tabcolsep}

\renewcommand{\arraystretch}{1.0}

\footnotesize 

\begin{longtable}{@{\extracolsep\fill}llllllll@{}}
\caption{Discovery of actinium, thorium, protactinium, and uranium isotopes. See page\ \pageref{tbl1te} for Explanation of Tables}
Isotope & First Author & Journal & Ref. & Method & Laboratory & Country & Year\\
\hline\\
\endfirsthead\\
\caption[]{(continued)}
Isotope & First author & Journal & Ref. & Method & Laboratory & Country & Year\\
\hline\\
\endhead
$^{206}$Ac & K. Eskola & Phys. Rev. C &\cite{1998Esk01}& FE & Jyv\"askyl\"a & Finland &1998 \\
$^{207}$Ac & M. Leino & Z. Phys. A &\cite{1994Lei01}& FE & Jyv\"askyl\"a & Finland &1994 \\
$^{208}$Ac & M. Leino & Z. Phys. A &\cite{1994Lei01}& FE & Jyv\"askyl\"a & Finland &1994 \\
$^{209}$Ac & K. Valli & Phys. Rev. &\cite{1968Val01}& FE & Berkeley & USA &1968 \\
$^{210}$Ac & K. Valli & Phys. Rev. &\cite{1968Val01}& FE & Berkeley & USA &1968 \\
$^{211}$Ac & K. Valli & Phys. Rev. &\cite{1968Val01}& FE & Berkeley & USA &1968 \\
$^{212}$Ac & K. Valli & Phys. Rev. &\cite{1968Val01}& FE & Berkeley & USA &1968 \\
$^{213}$Ac & K. Valli & Phys. Rev. &\cite{1968Val01}& FE & Berkeley & USA &1968 \\
$^{214}$Ac & K. Valli & Phys. Rev. &\cite{1968Val01}& FE & Berkeley & USA &1968 \\
$^{215}$Ac & K. Valli & Phys. Rev. &\cite{1968Val01}& FE & Berkeley & USA &1968 \\
$^{216}$Ac & H. Rotter & Sov. J. Nucl. Phys. &\cite{1967Rot01}& FE & Dubna & Russia &1967 \\
$^{217}$Ac & T. Nomura & Phys. Lett. B &\cite{1972Nom01}& FE & RIKEN & Japan &1972 \\
$^{218}$Ac & J. Borggreen & Phys. Rev. C &\cite{1970Bor01}& FE & Berkeley & USA &1970 \\
$^{219}$Ac & J. Borggreen & Phys. Rev. C &\cite{1970Bor01}& FE & Berkeley & USA &1970 \\
$^{220}$Ac & J. Borggreen & Phys. Rev. C &\cite{1970Bor01}& FE & Berkeley & USA &1970 \\
$^{221}$Ac & R.L. Hahn & Nucl. Phys. A &\cite{1968Hah01}& LP & Oak Ridge & USA &1968 \\
$^{222}$Ac & W.W. Meinke & Phys. Rev. &\cite{1949Mei01}& LP & Berkeley & USA &1949 \\
$^{223}$Ac & A. Ghiorso & Phys. Rev. &\cite{1948Ghi01}& LP & Berkeley & USA &1948 \\
$^{224}$Ac & A. Ghiorso & Phys. Rev. &\cite{1948Ghi01}& LP & Berkeley & USA &1948 \\
$^{225}$Ac & F. Hagemann & Phys. Rev. &\cite{1947Hag01}& RD & Argonne & USA &1947 \\
$^{226}$Ac & W.W. Meinke & Phys. Rev. &\cite{1950Mei01}& LP & Berkeley & USA &1950 \\
$^{227}$Ac & F. Giesel & Ber. Deuts. Chem. Ges. &\cite{1902Gie01}& RD & Braunschweig & Germany &1902 \\
$^{228}$Ac & O. Hahn & Phys. Z. &\cite{1908Hah01}& RD & Berlin & Germany &1908 \\
$^{229}$Ac & F. Depocas & Phys. Rev. &\cite{1952Dep01}& LP & Chalk River & Canada &1952 \\
$^{230}$Ac & K. Chayawattanangkur & J. Inorg. Nucl. Chem. &\cite{1973Cha01}& PN & Mainz & Germany &1973 \\
$^{231}$Ac & K. Chayawattanangkur & J. Inorg. Nucl. Chem. &\cite{1973Cha01}& PN & Mainz & Germany &1973 \\
$^{232}$Ac & K.-L. Gippert & Nucl. Phys. A &\cite{1986Gip01}& TR & Darmstadt & Germany &1986 \\
$^{233}$Ac & Y.Y. Chu & Phys. Rev. C &\cite{1983Chu01}& SP & Brookhaven & USA &1983 \\
$^{234}$Ac & K.-L. Gippert & Nucl. Phys. A &\cite{1986Gip01}& TR & Darmstadt & Germany &1986 \\
$^{235}$Ac & F. Bosch & Int. J. Mass Spectrom. &\cite{2006Bos01}& PF & Darmstadt & Germany &2006 \\
$^{236}$Ac & L. Chen & Phys. Lett. B &\cite{2010Che01}& PF & Darmstadt & Germany &2010 \\
 & & & & & &  \\
 & & & & & &  \\
$^{208}$Th & J.A. Heredia & Eur. Phys. J. A &\cite{2010Her01}& PF & Darmstadt & Germany &2010 \\
$^{209}$Th & H. Ikezoe & Phys. Rev. C &\cite{1996Ike01}& FE & JAERI & Japan &1996 \\
$^{210}$Th & J. Uusitalo & Phys. Rev. C &\cite{1995Uus01}& FE & Jyv\"askyl\"a & Finland &1995 \\
$^{211}$Th & J. Uusitalo & Phys. Rev. C &\cite{1995Uus01}& FE & Jyv\"askyl\"a & Finland &1995 \\
$^{212}$Th & D. Vermeulen & Z. Phys. A &\cite{1980Ver01}& FE & Darmstadt & Germany &1980 \\
$^{213}$Th & K. Valli & Phys. Rev. &\cite{1968Val02}& FE & Berkeley & USA &1968 \\
$^{214}$Th & K. Valli & Phys. Rev. &\cite{1968Val02}& FE & Berkeley & USA &1968 \\
$^{215}$Th & K. Valli & Phys. Rev. &\cite{1968Val02}& FE & Berkeley & USA &1968 \\
$^{216}$Th & K. Valli & Phys. Rev. &\cite{1968Val02}& FE & Berkeley & USA &1968 \\
$^{217}$Th & K. Valli & Phys. Rev. &\cite{1968Val02}& FE & Berkeley & USA &1968 \\
$^{218}$Th & K. Hiruta & Phys. Lett. B &\cite{1973Hir01}& FE & RIKEN & Japan &1973 \\
$^{219}$Th & O. Hausser & Phys. Rev. Lett. &\cite{1973Hau01}& FE & Chalk River & Canada &1973 \\
$^{220}$Th & O. Hausser & Phys. Rev. Lett. &\cite{1973Hau01}& FE & Chalk River & Canada &1973 \\
$^{221}$Th & D.F. Torgerson & Nucl. Phys. A &\cite{1970Tor01}& FE & Yale & USA &1970 \\
$^{222}$Th & D.F. Torgerson & Nucl. Phys. A &\cite{1970Tor01}& FE & Yale & USA &1970 \\
$^{223}$Th & W.W. Meinke & Phys. Rev. &\cite{1952Mei01}& LP & Berkeley & USA &1952 \\
$^{224}$Th & W.W. Meinke & Phys. Rev. &\cite{1949Mei01}& LP & Berkeley & USA &1949 \\
$^{225}$Th & W.W. Meinke & Phys. Rev. &\cite{1949Mei01}& LP & Berkeley & USA &1949 \\
$^{226}$Th & M.H. Studier & Phys. Rev. &\cite{1948Stu01}& LP & Argonne & USA &1948 \\
$^{227}$Th & O. Hahn & Nature &\cite{1906Hah01}& RD & McGill & Canada &1906 \\
$^{228}$Th & O. Hahn & Proc. Roy. Soc. A &\cite{1905Hah01}& RD & London & UK &1905 \\
$^{229}$Th & F. Hagemann & Phys. Rev. &\cite{1947Hag01}& RD & Argonne & USA &1947 \\
$^{230}$Th & B.B. Boltwood & Nature &\cite{1907Bol01}& RD & Yale & USA &1907 \\
$^{231}$Th & G.N. Antonoff & Phil. Mag. &\cite{1911Ant01}& RD & Manchester & UK &1911 \\
$^{232}$Th & G.C. Schmidt & Ann. Physik &\cite{1898Sch01}& RD & Erlangen& Germany &1898 \\
$^{233}$Th & O. Hahn & Naturewiss. &\cite{1935Hah01}& NC & Berlin & Germany &1935 \\
$^{234}$Th & W. Crookes & Proc. Roy. Soc. A &\cite{1900Cro01}& RD & London & UK &1900 \\
$^{235}$Th & N. Trautmann & Radiochim. Acta &\cite{1969Tra01}& LP & Mainz & Germany &1969 \\
$^{236}$Th & N. Kaffrell & Z. Naturforsch. &\cite{1973Kaf01}& PN & Mainz & Germany &1973 \\
$^{237}$Th & S. Yuan & Z. Phys. A &\cite{1993Yua01}& LP & Lanzhou & China &1993 \\
$^{238}$Th & J. He & Phys. Rev. C &\cite{1999He01}& TR & Lanzhou & China &1999 \\
 & & & & & &  \\
 & & & & & &  \\
$^{212}$Pa & S. Mitsuoka & Phys. Rev. C &\cite{1997Mit01}& FE & JAERI & Japan &1997 \\
$^{213}$Pa & V. Ninov & Z. Phys. A &\cite{1995Nin01}& FE & Darmstadt & Germany &1995 \\
$^{214}$Pa & V. Ninov & Z. Phys. A &\cite{1995Nin01}& FE & Darmstadt & Germany &1995 \\
$^{215}$Pa & K.-H. Schmidt & Nucl. Phys. A &\cite{1979Sch01}& FE & Darmstadt & Germany &1979 \\
$^{216}$Pa & G.Ya. Sung-Ching-Yang & Sov. J. Nucl. Phys. &\cite{1972Sun01}& FE & Dubna & Russia &1972 \\
$^{217}$Pa & K. Valli & Phys. Rev. &\cite{1968Val02}& FE & Berkeley & USA &1968 \\
$^{218}$Pa & K.-H. Schmidt & Nucl. Phys. A &\cite{1979Sch01}& FE & Darmstadt & Germany &1979 \\
$^{219}$Pa & Z. Liu & Nucl. Instrum. Meth. A &\cite{2005Liu01}& PF & Darmstadt & Germany &2005 \\
$^{220}$Pa & Z. Liu & Nucl. Instrum. Meth. A &\cite{2005Liu01}& PF & Darmstadt & Germany &2005 \\
$^{221}$Pa & R. Hingmann & Z. Phys. A &\cite{1983Hin01}& FE & Darmstadt & Germany &1983 \\
$^{222}$Pa & J. Borggreen & Phys. Rev. C &\cite{1970Bor01}& FE & Berkeley & USA &1970 \\
$^{223}$Pa & J. Borggreen & Phys. Rev. C &\cite{1970Bor01}& FE & Berkeley & USA &1970 \\
$^{224}$Pa & P.A. Tove & Ark. Fysik &\cite{1958Tov01}& LP & Uppsala & Sweden &1958 \\
$^{225}$Pa & P.A. Tove & Ark. Fysik &\cite{1958Tov01}& LP & Uppsala & Sweden &1958 \\
$^{226}$Pa & W.W. Meinke & Phys. Rev. &\cite{1949Mei01}& LP & Berkeley & USA &1949 \\
$^{227}$Pa & A. Ghiorso & Phys. Rev. &\cite{1948Ghi01}& LP & Berkeley & USA &1948 \\
$^{228}$Pa & A. Ghiorso & Phys. Rev. &\cite{1948Ghi01}& LP & Berkeley & USA &1948 \\
$^{229}$Pa & E.K. Hyde & Nat. Nucl. Ener. Ser. &\cite{1949Hyd02}& LP & Berkeley & USA &1949 \\
$^{230}$Pa & M.H. Studier & Phys. Rev. &\cite{1948Stu01}& LP & Argonne & USA &1948 \\
$^{231}$Pa & O. Hahn & Phys. Z. &\cite{1918Hah01}& RD & Berlin & Germany &1918 \\
$^{232}$Pa & J.W. Gofman & Nat. Nucl. Ener. Ser. &\cite{1949Gof01}& LP & Berkeley & USA &1949 \\
$^{233}$Pa & L. Meitner & Z. Phys. &\cite{1938Mei01}& NC & Berlin & Germany &1938 \\
$^{234}$Pa & K. Fajans & Naturwiss. &\cite{1913Faj01}& RD & Karlsruhe & Germany &1913 \\
$^{235}$Pa & W.W. Meinke & Phys. Rev. &\cite{1950Mei01}& LP & Berkeley & USA &1950 \\
$^{236}$Pa & G. Wolzak & Radiochim. Acta &\cite{1963Wol01}& LP & Amsterdam & Netherlands &1963 \\
$^{237}$Pa & W.W.T. Crane & Phys. Rev. &\cite{1954Cra01}& LP & Berkeley & USA &1954 \\
$^{238}$Pa & N. Trautmann & Z. Naturforsch. &\cite{1968Tra01}& LP & Mainz & Germany &1968 \\
$^{239}$Pa & S. Yuan & Z. Phys. A &\cite{1995Yua01}& TR & Lanzhou & China &1995 \\
 & & & & & &  \\
 & & & & & &  \\
$^{217}$U & O.N. Malyshev & Eur. Phys. J. A &\cite{2000Mal01}& FE & Dubna & Russia &2000 \\
$^{218}$U & A.N. Andreyev & Z. Phys. A &\cite{1992And01}& FE & Dubna & Russia &1992 \\
$^{219}$U & A.N. Andreyev & Z. Phys. A &\cite{1993And01}& FE & Dubna & Russia &1993 \\
$^{220}$U & & & & & & & \\
$^{221}$U & & & & & & & \\
$^{222}$U & R. Hingmann & Z. Phys. A &\cite{1983Hin01}& FE & Darmstadt & Germany &1983 \\
$^{223}$U & A.N. Andreyev & Sov. J. Nucl. Phys. &\cite{1991And01}& FE & Dubna & Russia &1991 \\
$^{224}$U & A.N. Andreyev & Sov. J. Nucl. Phys. &\cite{1991And01}& FE & Dubna & Russia &1991 \\
$^{225}$U & A.N. Andreyev & Sov. J. Nucl. Phys. &\cite{1989And01}& FE & Dubna & Russia &1989 \\
$^{226}$U & V.E. Viola & Nucl. Phys. A &\cite{1973Vio01}& LP & Maryland & USA &1973 \\
$^{227}$U & W.W. Meinke & Phys. Rev. &\cite{1952Mei01}& LP & Berkeley & USA &1952 \\
$^{228}$U & W.W. Meinke & Phys. Rev. &\cite{1949Mei01}& LP & Berkeley & USA &1949 \\
$^{229}$U & W.W. Meinke & Phys. Rev. &\cite{1949Mei01}& LP & Berkeley & USA &1949 \\
$^{230}$U & M.H. Studier & Phys. Rev. &\cite{1948Stu01}& LP & Argonne & USA &1948 \\
$^{231}$U & D.W. Osborne & Nat. Nucl. Ener. Ser. &\cite{1949Osb01}& LP & Berkeley & USA &1949 \\
$^{232}$U & J.W. Gofman & Nat. Nucl. Ener. Ser. &\cite{1949Gof01}& LP & Berkeley & USA &1949 \\
$^{233}$U & G.T. Seaborg & Phys. Rev. &\cite{1947Sea01}& NC & Berkeley & USA &1947 \\
$^{234}$U & H. Geiger & Phil. Mag. &\cite{1912Gei01}& RD & Manchester & UK &1912 \\
$^{235}$U & A.J. Dempster & Nature &\cite{1935Dem03}& MS & Chicago & USA &1935 \\
$^{236}$U & A. Ghiorso & Phys. Rev. &\cite{1951Ghi01}& NC & Argonne & USA &1951 \\
$^{237}$U & Y. Nishina & Phys. Rev. &\cite{1940Nis01}& NC & Tokyo & Japan &1940 \\
$^{238}$U & H. Becquerel & Compt. Rend. Acad. Sci. &\cite{1896Bec01}& RD & Paris & France &1896 \\
$^{239}$U & L. Meitner & Z. Phys. &\cite{1937Mei01}& NC & Berlin & Germany &1937 \\
$^{240}$U & J.D. Knight & Phys. Rev. &\cite{1953Kni01}& NC & Los Alamos & USA &1953 \\
$^{241}$U & & & & & & & \\
$^{242}$U & P.E. Haustein & Phys. Rev. C &\cite{1979Hau01}& LP & Brookhaven & USA &1979 \\
 & & & & & &  \\
 & & & & & &  \\
\end{longtable}

\end{document}